\documentclass[twoside,twocolumn,9pt]{article}
\usepackage{extsizes}
\usepackage[super,sort&compress,comma]{natbib} 
\usepackage[version=3]{mhchem}
\usepackage[left=1.5cm, right=1.5cm, top=1.785cm, bottom=2.0cm]{geometry}
\usepackage{balance}
\usepackage{mathptmx}
\usepackage{sectsty}
\usepackage{graphicx} 
\usepackage{dsfont}
\usepackage{lastpage}
\usepackage{multirow}
\usepackage{amssymb}
\usepackage[format=plain,justification=justified,singlelinecheck=false,font={stretch=1.125,small,sf},labelfont=bf,labelsep=space]{caption}
\usepackage{float}
\usepackage{fancyhdr}
\usepackage{fnpos}
\usepackage[english]{babel}
\addto{\captionsenglish}{%
  
}
\DeclareRobustCommand*{\citen }[1]{
  \begingroup
    \romannumeral-`\x
    \setcitestyle{numbers}
    \cite{#1}
  \endgroup   
}
\usepackage{array}
\usepackage{droidsans}
\usepackage{charter}
\usepackage[T1]{fontenc}
\usepackage[dvipsnames]{xcolor}
\usepackage{setspace}
\usepackage[compact]{titlesec}
\usepackage{hyperref}
\usepackage{booktabs} 

\usepackage[textsize=tiny]{todonotes}

\usepackage{epstopdf}

\definecolor{cream}{RGB}{222,217,201}
\newcommand{\rev}[1]{\textcolor{black}{#1}}

\begin{document}

\pagestyle{fancy}
\thispagestyle{plain}
\fancypagestyle{plain}{
\renewcommand{\headrulewidth}{0pt}
}

\makeFNbottom
\makeatletter
\renewcommand\LARGE{\@setfontsize\LARGE{15pt}{17}}
\renewcommand\Large{\@setfontsize\Large{12pt}{14}}
\renewcommand\large{\@setfontsize\large{10pt}{12}}
\renewcommand\footnotesize{\@setfontsize\footnotesize{7pt}{10}}
\makeatother

\renewcommand{\thefootnote}{\fnsymbol{footnote}}
\renewcommand\footnoterule{\vspace*{1pt}%
\color{cream}\hrule width 3.5in height 0.4pt \color{black}\vspace*{5pt}} 
\setcounter{secnumdepth}{5}

\makeatletter 
\renewcommand\@biblabel[1]{#1}            
\renewcommand\@makefntext[1]%
{\noindent\makebox[0pt][r]{\@thefnmark\,}#1}
\makeatother 
\renewcommand{\figurename}{\small{Fig.}~}
\sectionfont{\sffamily\Large}
\subsectionfont{\normalsize}
\subsubsectionfont{\bf}
\setstretch{1.125} 
\setlength{\skip\footins}{0.8cm}
\setlength{\footnotesep}{0.25cm}
\setlength{\jot}{10pt}
\titlespacing*{\section}{0pt}{4pt}{4pt}
\titlespacing*{\subsection}{0pt}{15pt}{1pt}

\fancyfoot{}
\fancyfoot[LO,RE]{\vspace{-7.1pt}Working Draft}
\fancyfoot[CO]{\vspace{-7.1pt}\hspace{13.2cm}}
\fancyfoot[CE]{\vspace{-7.2pt}\hspace{-14.2cm}}
\fancyfoot[RO]{\footnotesize{\sffamily{1--\pageref{LastPage} ~\textbar  \hspace{2pt}\thepage}}}
\fancyfoot[LE]{\footnotesize{\sffamily{\thepage~\textbar\hspace{3.45cm} 1--\pageref{LastPage}}}}
\fancyhead{}
\renewcommand{\headrulewidth}{0pt} 
\renewcommand{\footrulewidth}{0pt}
\setlength{\arrayrulewidth}{1pt}
\setlength{\columnsep}{6.5mm}
\setlength\bibsep{1pt}

\makeatletter 
\newlength{\figrulesep} 
\setlength{\figrulesep}{0.5\textfloatsep} 

\newcommand{\topfigrule}{\vspace*{-1pt}%
\noindent{\color{cream}\rule[-\figrulesep]{\columnwidth}{1.5pt}} }

\newcommand{\botfigrule}{\vspace*{-2pt}%
\noindent{\color{cream}\rule[\figrulesep]{\columnwidth}{1.5pt}} }

\newcommand{\dblfigrule}{\vspace*{-1pt}%
\noindent{\color{cream}\rule[-\figrulesep]{\textwidth}{1.5pt}} }

\makeatother

\twocolumn[
  \begin{@twocolumnfalse}
  
\vspace{1em}
\sffamily

\noindent \LARGE{\textbf{Role of interaction anisotropy in polymer cononsolvency: insights from the Flory-Huggins-Potts framework}} \\

\noindent\large{Satyen Dhamankar and Michael A. Webb$^*$} \\
 \noindent{Department of Chemical and Biological Engineering, Princeton University, Princeton, NJ 08544} \\

\noindent\normalsize{Cononsolvency occurs when mixing two good solvents creates poor-solvent conditions for polymers over specific composition ranges, causing macroscopic phase separation or microscopic chain collapse. Despite its technological and biophysical relevance, the connection between macroscopic and microscopic manifestations of cononsolvency remains unclear. A key challenge is identifying which interactions govern cononsolvency: coarse-grained analyses like standard Flory--Huggins models assume purely isotropic interactions, while atomistic simulations contain complex anisotropic interactions that cannot be precisely controlled or isolated. Here, we address the role of interaction anisotropy using the Flory-Huggins-Potts framework, which yields $\chi$ as a thermodynamic average over both configurational and internal-state coarse-grained degrees of freedom. This enables controlled comparison between systems with isotropic versus orientation-dependent interactions that share identical effective $\chi$ parameters, either driving cononsolvency by strong solvent-cosolvent affinity or preferential polymer-cosolvent affinity. While pairs of systems exhibit equivalent macroscopic phase behavior, lattice Monte Carlo simulations reveal that those featuring orientation-dependent interactions generate distinct collapse signatures, particularly in reentrant coil-globule transitions or characteristics of the solvation structure. These results demonstrate how microscopic interactions influences cononsolvency behavior beyond what effective $\chi$ parameters alone predict. 
} \\ 

\renewcommand*\rmdefault{bch}\normalfont\upshape
\rmfamily

\vspace{0.6cm}
 \end{@twocolumnfalse} \vspace{0.6cm}

  ]

\renewcommand*\rmdefault{bch}\normalfont\upshape
\rmfamily
\section*{}
\vspace{-1cm}

\section{Introduction}

Cononsolvency in polymer solutions arises when mixing two individually good solvents yields poor-solvent conditions for a polymer.  Under such conditions, at a macroscopic level, the solution may phase separate\cite{R:1968_Heskins_Solution,R:2022_Bharadwaj_Cononsolvency,R:2003_Takahashi_Effects,R:2017_Jia_Concentration,R:1980_Fernandez-Pierola_Co} or, at a microscopic level, the characteristic size of polymer chains may decrease.\cite{R:2012_Wang_Conformational,R:2001_Zhang_Water/Methanol,R:2022_Zhao_Effects,R:2021_Zhao_Proline,R:2010_Hao_Origin,R:2010_Sun_Role,R:2013_Hore_Co,R:2016_Jia_Re,R:2017_Budkov_Flory,R:2019_Pooch_Poly2,R:2019_Zuo_Water/Cosolvent,R:2019_Yong_Co} 
This composition-sensitive behavior has relevance across scientific and technological applications, such as detecting volatile organic compounds by measuring the response of a hydrogel,\cite{R:2019_Zhang_Volatile} measuring enantiomeric excess in organic synthesis by observing the fluorescent enhancement,\cite{R:2018_Nian_Racemic} altering rates of organic reactions at liquid-liquid interfaces,\cite{R:2020_Kleinschmidt_Microgel} tuning the frictional properties of polymer brushes by adjusting solution composition,\cite{R:2016_Yu_Tunable,R:2014_Chen_Collapse} and affecting the transport of molecules through a porous polymer matrix.\cite{R:2019_Wang_Multi‐Responsive,R:2021_Yong_Regulating, R:2016_Yu_Tunable} Furthermore, the physics governing cononsolvency has profound implications for understanding biological phase separation, protein denaturation,\cite{R:2019_Mills_Protein, R:2005_Valente_Second,R:2002_Xiao_Cosolvent,R:2020_Reddy_Cosolvent,R:2013_Canchi_Cosolvent,R:2023_Gazi_Conformational} and cell recovery in tissues.\cite{R:2010_Stuart_Emerging,R:2010_Nettles_Applications,R:2016_Rodriguez-Cabello_Elastin}
There is thus significant interest to understand polymer cononsolvency and related phenomena.

Multiple mechanisms of polymer cononsolvency have been proposed on the basis of theoretical, computational, and experimental analyses. \rev{The preferential mixing mechanism attributes polymer collapse to favorable solvent–cosolvent interactions, where the polymer sacrifices conformational entropy to promote solvent–cosolvent mixing.\cite{R:2016_Jia_Re,R:2017_Jia_Concentration,R:2019_Zuo_Water/Cosolvent,R:2015_Dudowicz_Theory,R:2017_Wang_Preferential,R:2019_Bharadwaj_Does} 
The seminal experimental study by Schild et al.\cite{R:1991_Schild_Cononsolvency} demonstrated, however, that perturbing water–methanol interactions (\(\chi_{12}\)) is insufficient to predict cononsolvency of PNIPAM (poly(N-isopropylacrylamide)) in a methanol–water solution.}
An alternative mechanism involves preferential polymer–cosolvent adsorption, where trace amounts of cosolvent selectively solvate the polymer and form “enthalpic bridges’’ between distal chain segments.\cite{R:2015_Mukherji_Co}  
These interactions can trigger phase separation or single-chain collapse.\cite{R:1985_Tanaka_polymer,R:1982_Tanaka_Collapse,R:2008_Tanaka_Temperature,R:2011_Tanaka_Preferential,R:2019_Dalgicdir_Improved,R:2019_Perez-Ramirez_P,R:2017_Budkov_Flory,R:2018_Budkov_Models,R:2017_Budkov_Statistical}  
Effects akin to preferential adsorption can also arise from entropic factors, such as size asymmetry between solvent and cosolvent leading to effective depletion interactions.\cite{R:2019_Bharadwaj_Does}
In another scenario, the cosolvent can act like a surfactant, such that polymer contraction allows the cosolvent to interact favorably with both solvent and polymer.\cite{R:2020_Bharadwaj_cosolvent}  
The concept of geometric frustration has also been used to explain cononsolvency, proposing that competition between cosolvent and solvent for polymer solvation destabilizes the local environment.\cite{R:2016_Pica_alternative,R:2017_Dalgicdir_Computational}  
Overall, these studies highlight polymer cononsolvency as a phenomenon with various competing physicochemical interactions.

 \rev{Previous studies using Flory-Huggins (FH) theory have demonstrated that phase separation can be induced when either solvent-cosolvent or polymer-solvent interactions dominate, even when the polymer is miscible with each solvent individually.\cite{R:2015_Dudowicz_Theory,R:2015_Dudowicz_Communication:,R:2024_Zhang_Phase} Furthermore, analyses of structure factors obtained following application of the random phase approximation provide similar conclusions. \cite{R:2020_Zhang_unified} 
    Nevertheless, mean-field insights into the energetic factors driving phase separation may not always translate into understanding microscopic physics, such as single-chain conformational behavior.\cite{R:2015_Mukherji_Co,R:2025_Oliver_When} 
    In this direction, field theoretic methods\cite{R:2017_Budkov_Flory,R:2017_Budkov_Statistical} have demonstrated that single-chain collapse indeed occurs given strong, preferential affinity of the polymer for one of the solvents, but there is no apparent coil-globule transition driven purely by solvent-cosolvent affinity.
    Recent developments in variational field-theoretic treatments\cite{R:2024_Liu_Variational} may enable further insights into single-chain behavior under such conditions.}

Molecular dynamics (MD) simulations of systems with generic chemical attributes (e.g., bead-spring polymers in monomeric solvents) have  effectively illustrated many of the aforementioned mechanisms of cononsolvency at a microscopic level.\cite{R:2019_Bharadwaj_Does,R:2015_Mukherji_Co,R:2021_Bharadwaj_interplay} A typical observation from such simulations is that the $R_\text{g}$ for a single polymer chain first decreases and then increases as cosolvent is progressively added. This scenario has been  induced by solvent-cosolvent mixing, preferential polymer-solvent adsorption, and the surfactant-like action of the cosolvent by analyzing the molecular environment of the polymer and carefully evaluating molecular affinities. \cite{R:2015_Mukherji_Co,R:2014_Mukherji_Polymer,R:2017_Sommer_AdsorptionAttraction,R:2018_Sommer_Gluonic,R:2017_Dalgicdir_Computational} 

\rev{MD simulations have also been used to examine the behavior of chemically specific systems, such as PNIPAM in water and ethanol.
In such cases, single-chain collapse, consistent with cononsolvency, has been observed.
Despite significant interest and detailed study, there is debate as to what precise molecular interactions or combination thereof begets this behavior.\cite{R:2013_Mukherji_Coil–Globule–Coil,R:2016_Mukherji_Relating,R:2017_Pica_Comment,R:2017_Vegt_Comment,R:2017_Mukherji_Reply} This highlights how microscopic drivers of cononsolvency can be non-trivial to resolve, even if the basic physics are well understood. }
\rev{This study does not adjudicate the specific atomistic interactions driving cononsolvency--it disentangles how systems with the same effective $\chi$ parameters show distinct molecular behavior.}

Several recent works have established quantitative connections between microscopic chain behavior and macroscopic phase separation in cononsolvency for systems with isotropic interactions. 
Zhang has comprehensively mapped the energetic conditions (i.e., $\chi$ parameters) that yield cononsolvency through mean-field analysis.\cite{R:2024_Zhang_Phase,R:2025_Zhang_Phase}
Previously, Zhang et al. also demonstrated via Wang-Landau simulations that trace cosolvent addition induces continuous coil-globule-coil transitions without requiring explicit solvent-solvent attraction.\cite{R:2022_Zhang_Conformation} 
By a complmenetary and distinct approach, Marcato et al. mapped  lattice models onto $\mathcal{O}(n)$-vector spin models to derive exact partition functions and field-theoretic descriptions amenable to analytics.\cite{R:2024_Marcato_Theory} 
Li et al. have established connections between single-chain collapse, multi-chain aggregation, and mean-field theory for block copolymers and homopolymer chain(s) in binary solvents, in the context of FH models\cite{R:2025_Li_novel} 
While these studies establish that specific combinations of scalar $\chi$ parameters derived from isotropic interactions produce cononsolvency, this leaves unresolved whether \rev{orientation-dependent interactions} fundamentally alters cononsolvency mechanisms or merely modulates existing ones.

While simple FH theory captures cononsolvency phenomena,
the apparent connection between thermoresponsive polymer solutions and cononsolvency\cite{R:2008_Tanaka_Temperature,R:2014_Scherzinger_Cononsolvency,R:2022_Bharadwaj_Cononsolvency} and highlighted importance of hydrogen bonding,\cite{R:2022_Yong_Cononsolvency,R:2008_Tanaka_Temperature} 
ultimately suggest that \rev{orientation-dependent interactions} may be relevant for at least some specific cononsolvency mechanisms.
Here, we directly investigate how \rev{anisotropic interactions} influences polymer cononsolvency in the context of a Flory-Huggins-Potts (FHP) framework,\cite{R:2024_Dhamankar_Asymmetry} 
which previously demonstrated that including orientation-dependent interactions within FH-like models enabled description of thermoresponsive phenomena (e.g., miscibility loops and heating-induced coil-globule transitions) without temperature-dependent $\chi$ parameters.
\rev{This work does not resolve mechanisms or debate for any specific system; instead, it clarifies how orientation-dependent interactions, like hydrogen bonding or even packing effects, influence cononsolvency phenomena within a single, tractable framework.
In particular, we demonstrate how systems with  the same effective $\chi$ interaction parameters, and thus identical macroscopic phase behavior, may exhibit different microscopic signatures of cononsolvency.}

\rev{The remainder of the paper is organized as follows. 
After presentation of methods,} we first identify established phase-separation regimes driven by solvent-cosolvent mixing versus polymer-cosolvent adsorption.
Then, we compare corresponding coil-globule-coil transitions across systems with identical effective $\chi$ parameters but different underlying energetic contributions using lattice Monte Carlo simulations.
Subsequent analysis reveals how anisotropic interactions generate microscopically distinct collapse pathways compared to isotropic systems, despite yielding identical macroscopic $\chi$ parameters. 
\rev{These findings provide additional insights into cononsolvency, by ascertaining the influence of additional degrees of freedom imparted by the FHP framework, and point towards strategies for tuning polymer response based on molecular-level interactions.  }


\section{Methods} 

\subsection{Flory-Huggins-Potts framework and stability analysis}

To enable description of orientation-dependent interactions,
we extend the recently developed FHP framework\cite{R:2024_Dhamankar_Asymmetry} to model ternary polymer solutions. 
In effect, the FHP framework augments conventional FH theory by assigning internal states to particles that modulate pairwise interactions. These states may be interpreted, for example, as capturing aspects of molecular orientation at a coarse-grained resolution, enabling expression of orientation-dependent interactions, whereas traditional FH assumes isotropic interactions.
While the motivation and detailed development of the FHP framework can be found in Ref.\citen{R:2024_Dhamankar_Asymmetry}, we briefly review some salient aspects. 

The behavior of the FHP model is governed by a well-defined Hamiltonian. We consider a fully occupied lattice of polymer, solvent, and cosolvent sites, with the system energy given by bonded and non-bonded contributions:
\begin{equation}\label{eq:hamiltonian}
    \mathcal{H} = \frac{1}{2}\sum_{i=1}^n \sum_{j \in \mathcal{N}(i)} \epsilon(\alpha_i, \alpha_j, \hat{\sigma}_i, \hat{\sigma}_j) + \sum_{k=1}^{N_\text{p}} \sum_{l=1}^{N_\text{m}-1} V\left(\vec{r}_{l+1}^{(k)}, \vec{r}_l^{(k)} \right),
\end{equation}
where $\alpha_i$ is the species type (monomer `m', solvent `s', or cosolvent `c') and $\hat{\sigma}_i$ is the unit vector representing the orientation of the particle at site $i$, and $\mathcal{N}(i)$ is the set of particles with which $i$ interacts. The function $\epsilon(\cdot)$ defines pairwise interaction energies, $n$ is the total number of lattice sites, $N_\text{p}$ is the number of polymer chains, $N_\text{m}$ is the number of monomers per chain, and $\vec{r}_l^{(k)}$ is the position of the $l$th monomer in chain $k$. The potential $V(\cdot)$ ensures bonded monomers remain within $\mathcal{N}(i)$.
The non-bonded interaction term is defined as:
\begin{align}\label{eq:expansion}
\epsilon(\alpha_i, \alpha_j, \hat{\sigma}_i, \hat{\sigma}_j) = \epsilon_{\alpha_i \alpha_j}^{\nparallel} + \Lambda(i,j) \Delta_{\alpha_i \alpha_j},
\end{align}
where $\Delta_{\alpha_i \alpha_j} = \epsilon_{\alpha_i \alpha_j}^{\parallel} - \epsilon_{\alpha_i \alpha_j}^{\nparallel}$ captures the difference between aligned and misaligned interactions, and $\Lambda(i,j) \in [0,1]$ distinguishes the degree of alignment. For convenience, we assume aligned interactions are stronger, so $\Delta_{\alpha_i \alpha_j} \leq 0$. When $\Lambda(i,j) $ or $\Delta_{\alpha_i \alpha_j} =0$, interactions are isotropic, and FHP reduces to standard FH theory.

We investigate $\Lambda(i,j)$ with a form given by
\begin{align}\label{eq:corr}
    \Lambda_{\text{corr}}(i, j) = \Theta(\hat{\sigma}_i \cdot \hat{\sigma}_j - \delta)
\end{align}
where $\Theta(\cdot)$ is the Heaviside step function, and $\delta$ defines the angular tolerance for alignment.
We refer to this interaction style as a ``correlation network,'' and it resembles interactions in Maier-Saupe theory.\cite{R:2022_Schoot_Molecular}
Correlation networks allow all neighbors to align cooperatively, since the function becomes nonzero when two vectors point nominally in the same direction. 
\rev{Importantly, the orientation vectors here are coarse-grained internal state variables that modulate local interaction energies. 
In one manifestation, these vectors could encode differences in molecular orientation, where the positioning of certain functional groups would influence the manner of interaction with other moieties. However, these vectors may also represent other forms of collective ordering, such as that seen in nematic liquid crystals or in hydrogen-bond networks in water, where many simultaneous, orientation-dependent interactions occur.}
\rev{While $\Delta_{\alpha_i \alpha_j}$ terms are not treated in chemically specific terms here, surveying its influence over different values phenomenologically probes the role of orientation-dependent interactions, irrespective of their origin.}

We note that the transition from simple FH to FHP introduces several additional parameters governing particle interactions. However, not all parameters are essential for modeling specific phenomena, and FHP parameters have been successfully fitted to experimental data with high precision.\cite{R:2024_Dhamankar_Asymmetry} Regularization techniques and physical constraints can yield parsimonious models when needed. Since our objective is to distinguish cononsolvency manifestations, \rev{we employ minimal models that introduce vary only two while keeping other variables fixed.}
\rev{In particular,} alignment-biased correlation network interactions are assigned between monomer–cosolvent and solvent–cosolvent pairs with fixed weights $p_v = 1.0$ and $p_\Omega = 0.25$ and the energetic penalty for misalignment is fixed $\Delta_{\text{mc}} = \Delta_{\text{sc}} = \Delta$; \rev{all other interactions are isotropic}. 
This minimal extension is sufficient to capture the anisotropic mechanisms discussed below while avoiding an unwieldy parameter space.
\rev{To provide some physical context, in the main text, we report results using    $\Delta = 0.5\epsilon_0$ where $\epsilon_0=k_\text{B}T$ is the unit of energy.
At 300 K, representative non‑covalent interactions span a wide energy range: water‑to‑amide hydrogen bonds are roughly –3 to –10 kcal/mol,\cite{R:1994_Dixon_Amide} $\pi-\pi$ contacts lie between $-3$ and $-5$ kcal/mol, \cite{R:2014_Huber_Heteroaromatic,R:2019_Molcanov_Towards} whereas water–alcohol hydrogen bonds are much weaker, about $-0.5$ kcal/mol.\cite{R:2015_Finneran_Hydrogen,R:2022_Swathi_Hydrogen} 
To bracket this spectrum, additional results for  $\Delta=-0.2\epsilon_0, -0.5 \epsilon_0, -0.8\epsilon _0$ are provided in the Supplementary Information.}

\subsection{Helmholtz energy by mean-field analysis}

Macroscopic phase behavior in the FHP framework is evaluated in the context of mean-field theory. All systems are considered incompressible, such that the total volume is given by $V = \sum_i n_i v_i$ where $n_i$ is the mole number and $v_i$ the molar volume of species $i$. If each polymer segment occupies one lattice site of volume $v_0$, then $v_\text{p} = N_\text{m} v_0$. The system composition is described by volume fractions $\phi_i = (n_i v_i)/V$, with $\sum_{i \in \{\text{p}, \text{s}, \text{c}\}} \phi_i = 1$. 
 
Following Ref.~\citen{R:2024_Dhamankar_Asymmetry}, but extended to three components, the intensive free energy of mixing per lattice site is:
\begin{equation}\label{eq:fmix}
\beta \Delta f_{\text{mix}} = \sum_i \frac{\phi_i}{(v_i/v_0)} \ln \phi_i + \frac{1}{2} \sum_j \sum_{k \neq j} \phi_j \phi_k\, \chi^{\text{FHP}}_{jk},
\end{equation}
where $\beta = (\text{k}_\text{B} T)^{-1}$ and $\chi^{\text{FHP}}_{ij}$ are FHP interaction parameters (with $\chi_{ii}^\text{FHP} = 0$). This expression mirrors the ternary FH form but with modified interaction terms:
\begin{equation}\label{eq:chieff}
\chi^{\text{FHP}}_{ij}(T) = \beta (z-2) \left( \epsilon_{ij}^{\nparallel} - \frac{1}{2}(\epsilon_{ii}^{\nparallel} + \epsilon_{jj}^{\nparallel}) \right) + \tilde{\chi}_{ij}(T),
\end{equation}
where the first term is essentially the FH $\chi$ at the misaligned energy scale, $z$ is the coordination number on the lattice, and $\tilde{\chi}_{ij}(T)$ is a perturbation term that accounts for the difference between aligned and misaligned interactions.
In particular, 
\begin{equation}\label{eq:chitilde}
\tilde{\chi}_{ij}(T) = \beta(z - 2) p_v \left( \tilde{\Delta}_{ij}(T) - \frac{1}{2} ( \tilde{\Delta}_{ii}(T) + \tilde{\Delta}_{jj}(T) ) \right)
\end{equation}
where $\tilde{\Delta}_{ij}(T)$ accounts for the free energy difference between aligned and misaligned states:
\begin{equation}\label{eq:correction}
\tilde{\Delta}_{ij}(T) = \frac{\Delta_{ij}}{1 + \left( \frac{1 - p_\Omega}{p_\Omega} \right) \exp( \Delta_{ij} / k_\text{B} T ) }.
\end{equation}
The terms $p_v$ and $p_\Omega$ are geometric factors related to what defines aligned interactions.
In particular, $p_v$ is the fraction of neighbors that can form aligned interactions, and $p_\Omega$ is the fraction of pairwise orientations that are classified as aligned for such neighbors.
\rev{While
eq. \ref{eq:fmix} retains the classical FH functional form dictated by symmetry and thermodynamic consistency, eqs. \ref{eq:chieff} and \ref{eq:chitilde} allow $\chi$-values to incorporate state-dependent interactions into the FH framework without modifying its structure. Thus, the key offering of leveraging the FHP conceptual framework is that the various terms that contribute to $\chi^{\text{FHP}}_{jk}$ are traceable to a well-defined and immutable microscopic Hamiltonian with controllable orientation-dependent energy terms.
}

For simplicity, all numerical results correspond to a simple cubic lattice where $\mathcal{N}(i)$ includes the nearest, next-nearest, and next-next-nearest neighbors of site $i$, yielding $z=26$. 
Furthermore, polymers are monodisperse with $N_\text{m}$ = 72, such that the volume occupied by a polymer chain is $v_\text{p} = N_\text{m}v_0$. The volume occupied by each solvent particle and cosolvent particle is $v_0$. Particle orientations are restricted to the 26 lattice directions.
\rev{Although choices regarding the lattice and the interaction neighborhood influences $z$, 
a different prescription is expected to only modify the energy scales at which phenomena are observed without qualitatively altering mechanisms.}

\subsection{Characterization of phase behavior}

The phase behavior of the system can be determined from eq.~\eqref{eq:fmix} by specifying the composition $\boldsymbol{\phi} = (\phi_\text{p}, \phi_\text{s}, \phi_\text{c})$, the molar volumes $\boldsymbol{v} = (v_\text{p}, v_\text{s}, v_\text{c})$, and the interaction parameters $\boldsymbol{X}^{\text{FHP}} = (\chi_{\text{ps}}^{\text{FHP}}, \chi_{\text{pc}}^{\text{FHP}}, \chi_{\text{sc}}^{\text{FHP}})$. 
Briefly, the stability of a homogeneous mixture is assessed by examining the sign of the determinant of the Hessian for the Helmholtz energy:
\begin{align}\label{eq:stability}
\mathbf{H} &= \begin{pmatrix}
    \frac{\partial ^2 \beta \Delta f_{\text{mix}} }{\partial \phi _{\text{p}}^2} & \frac{\partial ^2 \beta \Delta f_{\text{mix}} }{\partial\phi _{\text{p}} \partial \phi _{\text{s}} }\\
    \frac{\partial ^2 \beta \Delta f_{\text{mix}} }{\partial\phi _{\text{p}} \partial \phi _{\text{s}} } & \frac{\partial ^2 \beta \Delta f_{\text{mix}} }{\partial \phi _{\text{s}}^2}
\end{pmatrix} \\
|\mathbf{H}| &= \frac{\partial ^2 \beta \Delta f_{\text{mix}} }{\partial \phi _{\text{p}}^2}\cdot\frac{\partial ^2 \beta \Delta f_{\text{mix}} }{\partial \phi _{\text{s}}^2} - \left(\frac{\partial ^2 \beta \Delta f_{\text{mix}} }{\partial\phi _{\text{p}} \partial \phi _{\text{s}} }\right)^2.
\end{align}
The spinodal boundary is defined by the locus of points for which $|\mathbf{H}| =0$, while negative values indicate an unstable mixture that would undergo phase separation. 
Here, we use the condition $|\mathbf{H}| < 0$ over some composition range as the criterion for determining whether a system exhibits cononsolvency at a macroscopic level.

For select systems that meet the criterion above, their phase behavior is more precisely characterized by mapping binodal boundaries and determining the compositions of coexisting phases.
Binodal boundaries are computed by first identifying critical points using constraints involving third-order derivatives \cite{R:1949_Scott_Thermodynamics,R:1949_Tompa_Critical}:
\begin{equation}\label{eq:E1}
\frac{\partial |\mathbf{H}| }{\partial \phi _{\text{p}}}\cdot\frac{\partial ^2 \beta \Delta f_{\text{mix}} }{\partial \phi _{\text{s}}^2} - \frac{\partial |\mathbf{H}| }{\partial \phi _{\text{s}}}\cdot \frac{\partial ^2 \beta \Delta f_{\text{mix}} }{\partial\phi _{\text{p}} \partial \phi _{\text{s}}} = 0
\end{equation}
and
\begin{equation}\label{eq:E2}
\frac{\partial |\mathbf{H}| }{\partial \phi _{\text{s}}}\cdot\frac{\partial ^2 \beta \Delta f_{\text{mix}} }{\partial \phi _{\text{p}}^2} - \frac{\partial |\mathbf{H}| }{\partial \phi _{\text{p}}}\cdot \frac{\partial ^2 \beta \Delta f_{\text{mix}} }{\partial\phi _{\text{p}} \partial \phi _{\text{s}}} = 0.
\end{equation}
Following identification of a critical point, coexistence lines are extended using a recently proposed natural parameter continuation algorithm\cite{R:2025_Dhamankar_Accelerating} akin to Gibbs-Duhem integration. 
As this approach iteratively identifies different compositions with equivalent chemical potentials, it also facilitates the construction of tie lines to indicate the equilibrium compositions of the coexisting phases.
For additional details regarding this algorithm or the utilization of eqs. {\eqref{eq:stability}-\eqref{eq:E2}}, readers are referred to Ref.~{\citen{R:2025_Dhamankar_Accelerating}} and the references therein.

\subsection{Monte Carlo simulations}
To characterize cononsolvency at a microscopic level, we perform lattice Monte Carlo (MC) simulations.
The MC simulations directly implement specific manifestations of the Hamiltonian defined by eq. {\eqref{eq:hamiltonian}}.
These simulations are performed only for systems with parameters that exhibit mixture instability (Table 1).
All simulations feature a $34\times 34\times 34$ simple cubic lattice with periodic boundaries and unit edge length; lattice sites are occupied by a mixture of solvent and cosolvent particles and a single polymer chain ($N_\text{m}=32$).
For each manifestation of a Hamiltonian, simulations are performed over a set of cosolvent fractions that approximately (within the limits afforded by the discrete lattice) correspond to $x_\text{c} \in [0,1]$ in increments of 0.1.

Systems are initialized by placing polymer chains on lattice sites using a self-avoiding random walk. Cosolvent particles are then randomly distributed across the remaining sites to achieve the target mole fraction, and any unoccupied sites are filled with solvent particles. All particles are randomly assigned one of twenty six possible orientations with uniform probability.
Configurational sampling is performed using a variety of MC moves. For solvent and cosolvent particles, these include orientation exchanges, collective orientation perturbations, and particle swaps. Polymer moves include end-rotation, forward and backward reptation, and chain regrowth with orientation updates. Orientation updates are also applied to solvent particles in contact with the polymer and to a randomly selected subset of lattice particles. See Table S1 in the Supporting Information for a summary of all moves. All move types are attempted with equal probability. 

Each simulation consists of $10^8$ MC moves, with configurations sampled every $10^4$ moves. The first half of each simulation is used for equilibration and the second half for collecting production data (see Supporting Information, Fig. S3 and Fig. S4). For each condition (parameter set, composition, and temperature), thirty independent simulations are performed for statistical analysis and uncertainty quantification.
In addition, to compute certain properties for an ideal mixture, thirty independent simulations are also run under athermal conditions (i.e., all energy parameters except the bonding and implied excluded-volume interactions are set to zero). 

\rev{It is worth nothing that the MC simulations technically reflect results for a chain in the canonical ensemble. Consequently, it is conceivable that substantial partitioning of solvent species into the near-space volume occupied by the polymer could reduce the effective (co)solvent mole fractions in the ``bulk,'' and such behavior would depend on the finite size of the simulation cell.
Here, we verified that the number of solvent and cosolvent molecules coordinated within the polymer solvation shell remains small relative to the total particle count ($< 0.5\%$, see SI, Fig. S2), ensuring that bulk composition remains effectively constant, and this should be sufficient to effectively draw the connection between macroscopic thermodynamics and the dilute single-chain physics.
In the future,
it may be preferable to employ a constant chemical potential ensemble,\cite{R:2017_Budkov_Flory,R:2018_Budkov_Models} where the polymer gyration volume exchanges solvent and cosolvent with an external reservoir, though this may require additional considerations to be compatible with the Hamiltonians explored here.}

\subsection{Conformational analysis}

Polymer size is characterized by the radius of gyration $R_\text{g}$. For a polymer with $N_\text{m}$ monomers at positions $\mathbf{r}_i$, the $R_\text{g}$ is 
\begin{equation}
    R_\text{g} = \sqrt{\frac{1}{N_\text{m}}\sum_{i=1}^{N_\text{m}} (\mathbf{r}_i - \mathbf{r}_\text{com})^2},
\end{equation}
with the polymer center of mass defined per usual
\begin{equation}
    \mathbf{r}_\text{com} = \frac{1}{N_\text{m}}\sum_{i=1}^{N_\text{m}}\mathbf{r}_i.
\end{equation}
For consistent comparisons across conditions, we report a normalized radius of gyration, $\frac{\langle R_\text{g} \rangle}{\langle R_\text{g}\rangle_{\text{ath}}}$, where \( \langle R_\text{g} \rangle \) is the ensemble-averaged $R_\text{g}$ for a given system and \( \langle{R}_\text{g}\rangle_{\text{ath}} \) is a  reference given by an athermal simulation incorporating only excluded-volume interactions. For chains with \( N_\text{m}=72 \), we empirically find that $\langle{R}_\text{g}\rangle_{\text{ath}} = 6.15\pm 0.005$. For simplicity, we use $\langle R_\text{g}\rangle _\text{ath}=6.15$ in out calculations.  

\begin{figure*}[htb]
\centering
\includegraphics[keepaspectratio]{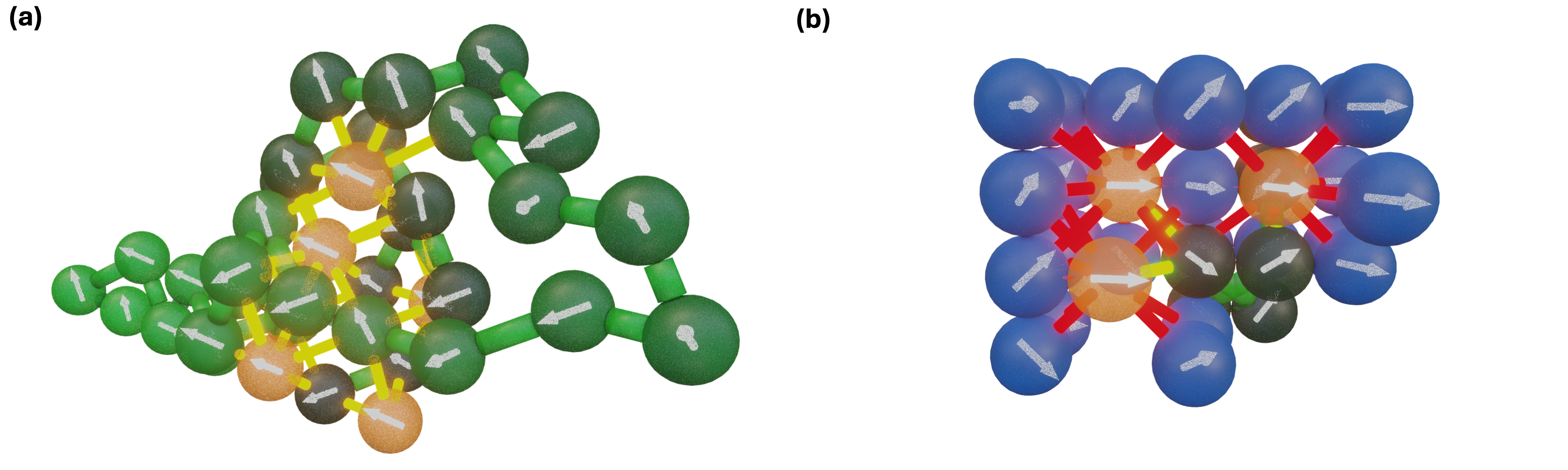}
\caption{Visualization of bridging and mediating cosolvents (orange) interacting with solvents (blue) and a polymer (green) in anisotropic $(\curlywedge)$ systems from Monte Carlo simulations. Monomer segments darken in shade along the contour of the polymer backbone. (a) Bridging cosolvents coordinating and aligning with distal monomer units (shown by gold bonds). (b) Mediating cosolvents coordinating aligned interactions with solvent particles (shown by red bonds) and with monomer units (shown by green-yellow bonds).}\label{fig:Fig1_molecular_renders}
\end{figure*}

\subsection{Solvation analysis}

To distinguish cononsolvency mechanisms at the single‑chain level, we compare the local solvation environment of the polymer from Monte Carlo simulations with ideal‑mixing predictions.  Deviations are quantified by the normalized excess number of interactions
\begin{align}
N_\kappa ^\text{ex} &= N_\kappa  - N_\kappa ^\text{id} \\ 
\frac{N_\kappa^\text{ex}}{N_\kappa^\text{id}} &= \tilde{N}_\kappa - 1
\end{align}
where $N_\kappa$ specifies the number of interactions under some restriction $\kappa$, and the superscripts `$\text{ex}$' and `$\text{id}$' indicate excess and ideal values, respectively.
The restriction of $\kappa$ for our analysis relates to some pairing of specific species or particles and/or whether interactions are aligned versus misaligned.
\rev{In the following, 
numbers of unique pairwise interactions between monomer, solvent, and cosolvent particles are represented with
$N_{\alpha_i \alpha_j}$ where $\alpha_i$ and $\alpha_j$ are the chemical species p, s, and c. } 

For ideal references, we first empirically determine $N^\text{id}_\text{mm}$ from MC simulation of an  athermal chain. Subsequently, the quantities $N_\text{ms}^\text{id}$ and $N_\text{mc}^\text{id}$ are inferred via
\begin{equation}\label{eq:id_Nms}
N_\text{ms}^\text{id} = (1-x_\text{c})N_{\text{ms} \bigcup \text{mc}}^{\text{id}},
\end{equation}
\begin{equation}\label{eq:id_Nmc}
N_\text{mc}^\text{id} = x_\text{c}N_{\text{ms} \bigcup \text{mc}}^{\text{id}},
\end{equation}
and
\begin{equation}\label{eq:id_Nms_mc}
N_{\text{ms} \bigcup \text{mc}}^{\text{id}} = zN_\text{m} - 2N_\text{mm}^\text{id},
\end{equation}
where the last expression for the number of either monomer-solvent or monomer-cosolvent interactions exploits the restriction that the overall number of interactions involving the polymer is conserved on the lattice. 
For chains with $N_\text{m} = 72$, we estimate $N_\text{mm}^\text{id}$ by computing the average number of monomer-monomer interactions from athermal simulations of a chain in solvent. 
These simulations yield  $\langle N_\text{mm}\rangle_\text{ath} = 150.12\pm 0.05$. 
For simplicity, we use  $N_\text{mm}^\text{id}=150$ in all calculations.  
Similarly, the expected number of solvent-cosolvent interactions is approximated as
\begin{equation}
    N_\text{sc}^\text{id} = zN_\text{s}  x_\text{c}.
\end{equation}


\rev{To assess the structure of the solvation shell, we compute the fraction of `bridging' cosolvents $\Phi _\text{b}^\curlywedge$, `mediating' cosolvent $\Phi _\text{m}^\curlywedge$, and `passive' cosolvents $\Phi _p^\curlywedge$ in the solvation shell.
The solvation shell is defined as the set of solvent and cosolvent particles in direct contact with the polymer \rev{i.e., nearest, next-nearest, or next-next-neareet neighbor)}. 
The classification of any given solvent as bridging, mediating, or passive depends on its underlying Hamiltonian. 
For systems with orientation-dependent interactions, a bridging cosolvent is is aligned and in contact with, two monomer units that lie at least three bonds apart, thereby forming an enthalpic bridge between distant chain segments (Fig.~\ref{fig:Fig1_molecular_renders}a). 
A mediating cosolvent  aligns with at least one monomer and one solvent particle but does not meet the bridging criterion. 
All other cosolvents are labeled passive (Fig.~\ref{fig:Fig1_molecular_renders}b). 
For systems with only isotropic interactions, we make analogous classifications based on the proximity and composition of the solvation shell, but the conditions regarding orienation are relaxed. 
In other words, a bridging cosolvent $\Phi_\text{b}^{\circ}$ contacts two monomers separated by at least three bonds; a mediating cosolvent $\Phi_\text{m}^{\circ}$ contacts one monomer and one solvent particle without bridging; and the remainder are passive $\Phi_\text{p}^{\circ}$. 
After classifying the various cosolvents,  we report the normalized mean energy per particle for bridging and mediating cosolvents, denoted $\bar{E}_b$ and $\bar{E}_m$, respectively; the passive set provides a baseline for comparison.}

\section{Results and discussion}

Our analysis proceeds in two parts. 
In Section {\ref{sec:phase}}, we use mean-field theory to identify the energetic regimes that drive cononsolvency through macroscopic phase separation. These results establish conceptual scaffolding and terminology consistent with prior literature. 
While similar results have appeared variously,\cite{R:2015_Dudowicz_Theory,R:2015_Dudowicz_Communication:,R:2025_Zhang_Phase,R:2024_Zhang_Phase,R:2025_Li_novel,R:2020_Zhang_unified}
we include this analysis for completeness since application of FHP is new, and specific results here directly inform our subsequent system selection. 
Readers familiar with cononsolvency may skim this section.
In Section {\ref{sec:sim}}, we employ lattice Monte Carlo simulations to examine microscopic behavior. We strategically select two pairs of systems with equivalent $\chi^\text{eff}_{ij}$ from phase-separating regimes, but for a given pair, system differ in their Hamiltonians; one features only isotropic interactions and the other incorporates orientation-dependent terms. 
Through detailed analysis of coil-globule-coil transitions, which are observed for all systems, we elucidate how interaction anisotropy influences cononsolvency at the microscopic level.


  

\begin{figure*}[htbp]
\centering
\includegraphics{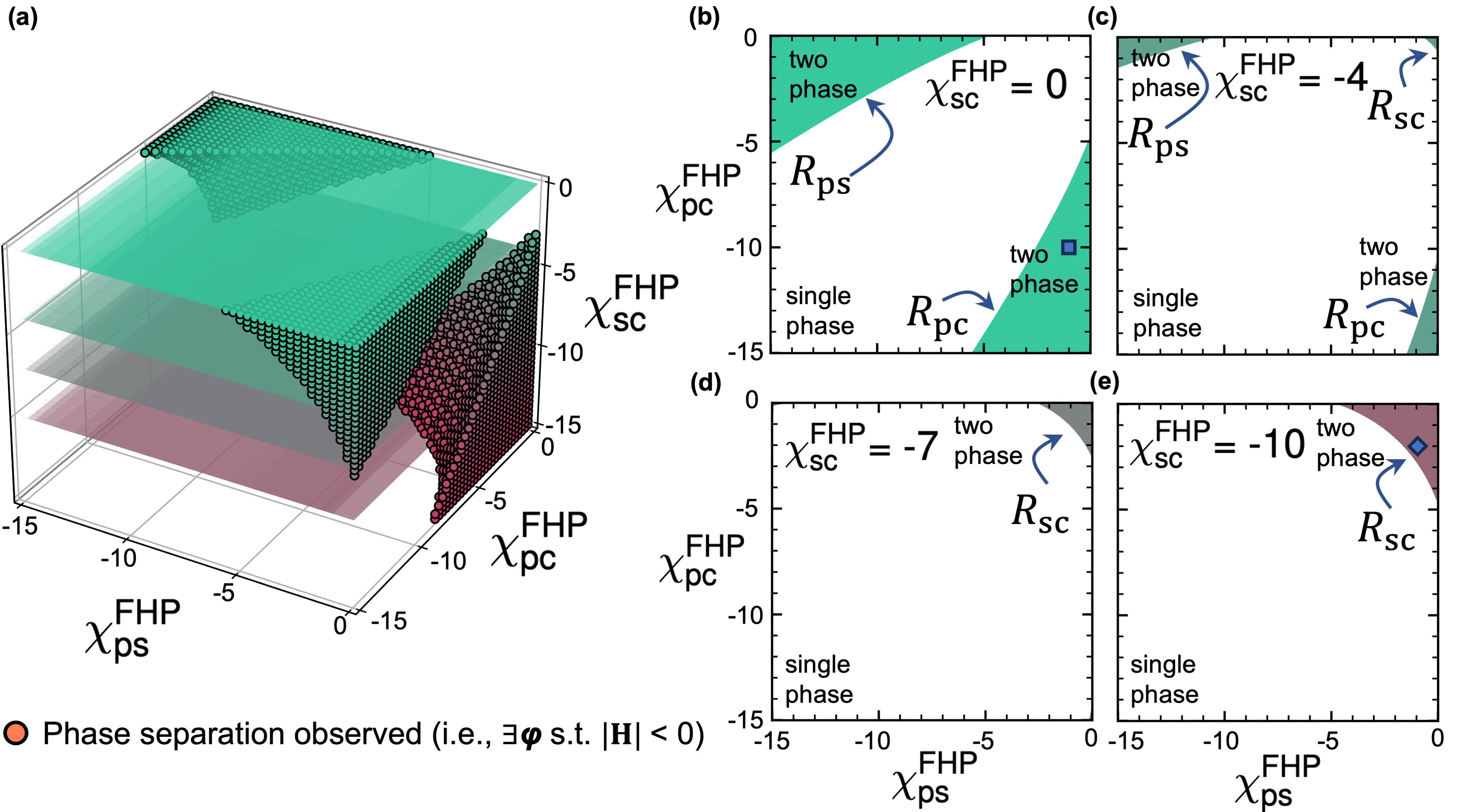}
\caption{Classification of phase behavior for ternary polymer solutions described by the FHP framework.  
(a) Regions in interaction parameter space $\boldsymbol{X}^\text{FHP} = (\chi^\text{FHP}_\text{ps}, \chi^\text{FHP}_\text{pc}, \chi^\text{FHP}_\text{sc})$ where phase separation is required at some composition $\boldsymbol{\phi}$, as determined by the mean-field stability criterion ($|\mathbf{H}| < 0$). Markers indicate parameter sets that yield instability; cross-sectional planes at $\chi^\text{FHP}_\text{sc} = 0$, $-4$, $-7$, and $-10$ correspond to panels (b-e).  
(b-e) Two-dimensional slices of (a) showing phase behavior as a function of $\chi^\text{FHP}_\text{ps}$ and $\chi^\text{FHP}_\text{pc}$ for fixed values of $\chi^\text{FHP}_\text{sc}$. Shaded regions indicate parameter combinations that guarantee phase separation at some composition. Regions of observed two-phase coexistence are labeled $R_\text{sc}$, $R_\text{ps}$, and $R_\text{pc}$, according to the most negative interaction parameter. Square and diamond markers in (b) and (e) denote parameter sets selected for further analysis in Section {\ref{sec:sim}}.
 }\label{fig:Fig2_ChiExploration}
\end{figure*}

\subsection{Analysis of phase behavior}\label{sec:phase}

To establish essential conceptual scaffolding and expectations surrounding cononsolvency,
we begin by identifying parameter regimes that display phase separation in ternary polymer solutions in the context of the FHP framework. 
In particular, we systematically explore the interaction-parameter space $\boldsymbol{X}^\text{FHP} = (\chi^\text{FHP}_\text{ps}, \chi^\text{FHP}_\text{pc}, \chi^\text{FHP}_\text{sc})$ and identify parameter sets for which $|\mathbf{H}| < 0$, signaling instability and thus phase separation, under the constraint that all components are mutually miscible ($\chi^\text{FHP}_{ij} < 0 \, \forall\, i,j$).   

Consistent with prior literature using FH theory,\cite{R:2015_Dudowicz_Theory,R:2015_Dudowicz_Communication:,R:2025_Zhang_Phase,R:2024_Zhang_Phase,R:2025_Li_novel}
there are three distinct parameter regimes characterized by the strongest interaction type present (Fig.~\ref{fig:Fig2_ChiExploration}). 
We denote these regimes as $R_{ij}$, corresponding to the region of parameter space where $(\chi^\text{FHP}_{ij} < \chi^\text{FHP}_{ik}, \chi^\text{FHP}_{jk} \leq 0)$ (i.e., the strongest attractive interaction is between species $i$ and $j$).
The emergence of cononsolvency, in terms of competing interactions, can be understood by examining successive cross sections in Fig.~\ref{fig:Fig2_ChiExploration}a.
Initially, at $\chi^\text{FHP}_\text{sc}=0$, two distinct regions, $R_\text{ps}$ and $R_\text{pc}$, exhibit phase separation for a broad range of parameters, for which polymer-solvent or polymer-cosolvent interactions are strongly favorable relative to interactions between solvent and cosolvent (Fig.~{\ref{fig:Fig2_ChiExploration}b). 
As solvent-cosolvent interactions become increasingly favorable (more negative $\chi^\text{FHP}_\text{sc}$), the areas of $R_\text{ps}$ and $R_\text{pc}$ shrink, as interactions are more balanced. 
Concurrently, a third region, $R_\text{sc}$, emerges where solvent-cosolvent interactions dominate. (Fig.~\ref{fig:Fig2_ChiExploration}c) 
Eventually, as solvent-cosolvent affinity strengthens further, $R_\text{ps}$ and $R_\text{pc}$ vanish entirely, leaving an expanded $R_\text{sc}$ region (Fig.~\ref{fig:Fig2_ChiExploration}d,e).
This analysis provides a baseline expectation that cononsolvency can be driven by different dominant interactions, be it polymer-solvent, polymer-cosolvent, or solvent-cosolvent, irrespective of microscopic details. 

Prior work has suggested that the shape, location, and extent of coexistence regions, as well as the orientation of tie lines, may reflect distinct underlying mechanisms of cononsolvency. \cite{R:2024_Zhang_Phase} 
To investigate this, we constructed full ternary phase diagrams for representative parameter sets drawn from each of the three regimes: $R_\text{ps}$, $R_\text{pc}$, and $R_\text{sc}$. 
Indeed, the shapes of binodal curves and the orientations of tie lines differ based on the dominant interaction present in the system (Supplementary Information, Fig. S1)
When interactions between a polymer and one of the solvents dominate, tie lines connect phases in which one is rich in polymer and the better solvent, while the other is rich in the lesser solvent and depleted in polymer. 
When solvent-cosolvent interactions dominate, the tie lines orient nearly perpendicular to the polymer composition axis, leading to coexistence between a very polymer-lean phase with mixed solvent and a polymer-rich phase.

\rev{Visually, the orientation of the tielines and the resultant coexistent composition suggest that there is a polymer-lean and polymer-rich phase. To verify that the reported instabilities correspond to cononsolvency, we performed an eigenmode analysis of the Hessian of the free energy at representative compositions within the binodal region. In the $\chi_\text{pc}$-dominated regime, the unstable mode corresponds to a channel where the polymer and cosolvent fluctuate together while the solvent fluctuates in the opposite direction. Conversely, in the $\chi_\text{sc}$-dominated regime, the instability primarily follows the polymer concentration while solvent and cosolvent fluctuate in phase. These results confirm that cononsolvency arises via distinct mechanisms in the two regimes. Full details of the eigenmode decomposition and channel classification are provided in the Supporting Information (Section S2).}

Nevertheless, while these observations may serve as useful indicators of the dominant underlying interactions, they remain qualitative and do not offer insight into the microscopic details of the interactions that underlie cononsolvency. 


\subsection{Analysis of simulations}\label{sec:sim}

The preceding results demonstrate how specific combinations of mean-field interaction parameters can induce macroscopic phase separation across a range of ternary compositions. We now turn to the question of whether such macroscopic behavior is echoed by a coil-globule transition induced by solvent-mixture composition and how this depends on whether interactions are isotropic or anisotropic in form.

To address this, we perform Monte Carlo (MC) simulations to study the conformational behavior and solvation-shell characteristics of a single polymer chain in mixed solvents for four systems. \rev{We present results for a chain with degree of polymerization $N_\text{m}=72$; corresponding data for a shorter chain with $N_\text{m}=32$ are provided in the SI (Figs. S9–S12). Both lengths show similar physics but the longer chain shows more pronounced effects.} In Section {\ref{sec:overview}}, we describe the selection of these systems and broadly categorize their behavior. 
The results supporting this categorization are then detailed in Sections {\ref{sec:sc}} and {\ref{sec:pc}}.

\subsubsection{Overview of systems and mechanisms}\label{sec:overview}

The four systems studied divide into two pairs based on their parameter sets (Table~\ref{tab:sim_parameters}).
One pair shares the same $\chi_{ij}^\text{FHP}$ parameter set from $R_\text{sc}$ (diamond marker in Fig. \ref{fig:Fig2_ChiExploration}e), and the other shares $\chi_{ij}^\text{FHP}$ from $R_\text{pc}$ (square marker in Fig.\ref{fig:Fig2_ChiExploration}b). 
For a given pair, systems differ in whether non-bonded interactions are orientation-independent (isotropic, denoted with `$\circ$') or orientation-dependent (anisotropic, denoted with `$\curlywedge$'). 
The complexity of anisotropy is restricted to only interactions involving cosolvent.
For simplicity, we impose that this is governed by a single energy scale, $\Delta_\text{$*$c} = \Delta_\text{mc} = \Delta_\text{sc}$.
In the main text, we focus on $\Delta_\text{$*$c} = 0.5$, while results for lesser and greater $\Delta_\text{$*$c}$ are provided in the Supplementary Information.
Furthermore,  parameters associated with monomer-monomer, solvent-solvent, cosolvent-cosolvent and monomer-solvent interactions are fixed across all systems ($\epsilon_{\text{mm}}^{\nparallel} = -1$, $\epsilon_{\text{ms}}^{\nparallel}  = -0.5416$, and $\epsilon_\text{ss}^\nparallel= \epsilon_\text{cc}^\nparallel=0$).

\begin{table}[h]
  \centering
  \caption{Parameters used in Monte Carlo simulations of FHP models. Parameters for label type $R_{ij}^\circ$ are for isotropic simulations $(\Delta _{ij}=0)$ and those for type $R_{ij}^\curlywedge$ are for anisotropic simulations $(\Delta _{ij}\neq 0)$. }\label{tab:sim_parameters}
  \setlength{\tabcolsep}{12pt}

  \renewcommand{\arraystretch}{1.6}
  
  \scriptsize
  \begin{tabular*}{\textwidth}{|c|ccc||cc|}
    \cline{1-6}
    \textbf{Label} & $\chi^\mathrm{FHP}_{\text{ps}}$ & $\chi^\mathrm{FHP}_{\text{pc}}$ & $\chi^\mathrm{FHP}_{\text{sc}}$ 
    & $\Delta_{\text{mc}}$ & $\Delta_{\text{sc}}$ \\
    \cline{1-6}
    $R_{\text{sc}}^\circ$
      & -1  & -2     & -10 
      & 0   & 0  \\
    \cline{1-6}
    $R_{\text{sc}}^\curlywedge$
      & -1        & -2      & -10 
      & -0.5   & -0.5\\
    \cline{1-6}\cline{1-6}\cline{1-6}
    $R_{\text{pc}}^\circ$
      & -1  & -10    & 0   
      & 0   & 0\\
    \cline{1-6}
    $R_{\text{pc}}^\curlywedge$
      & -1  & -10    & 0 
      & -0.5   & -0.5\\
      \cline{1-6}
  \end{tabular*}
\end{table}

In all systems, we find evidence of single-chain collapse at certain mixture compositions based on observed $R_\text{g}$ relative to that in either pure solvent. Based on further analysis of the solvation environment around the polymer, we then associate these chain collapses with specific mechanisms within the vernacular of the cononsolvency literature. 
Simulations with parameter sets that are derived from $R_{\text{sc}}$ show a collapse through preferential mixing of solvent and cosolvent; this mechanism tends to yield a ``dry'' globule.
In this case, including anisotropic interactions with cosolvent species qualitatively alters the cosolvent-induced coil-globule transition.
Parameter sets from $R_{\text{pc}}$ collapse through preferential adsorption of cosolvent; this mechanism tends to yield a ``wet'' globule.
For this situation, the role of anisotropy is apparent only in nuanced changes to the polymer solvation environment in its collapsed state.   
 
\begin{figure}[htbp]
    \centering
    \includegraphics[keepaspectratio]{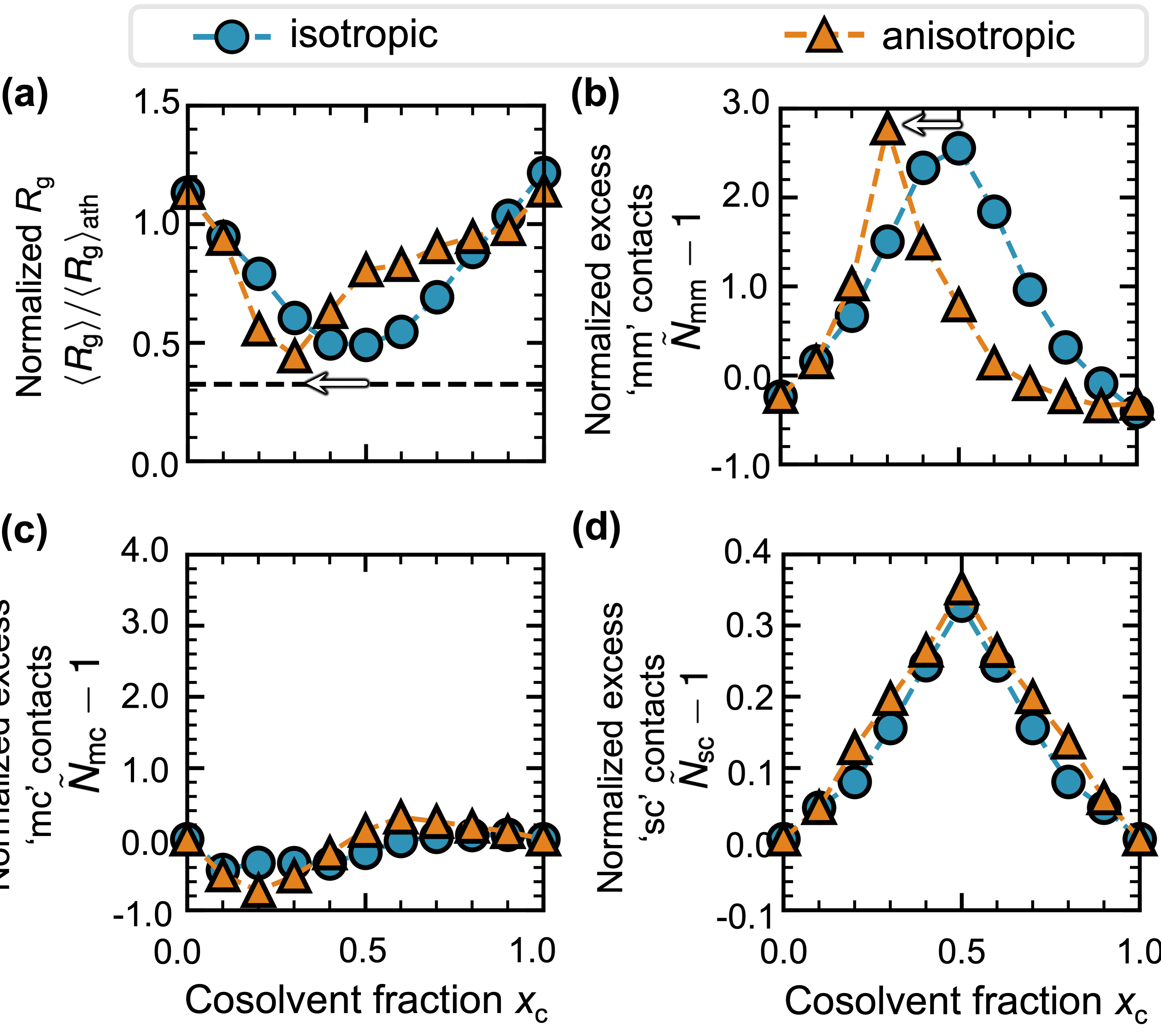}
    \caption{A comparison of conformational characteristics and environment of a polymer from $R_\text{sc}^\circ$ (blue circles) and $R_\text{sc}^\curlywedge$ (orange triangles). (a) Normalized single-chain $R_\text{g}$ and normalized excess (b) monomer-monomer, (c) monomer-cosolvent and (d) solvent-cosolvent interactions as cosolvent fraction $x_c$ is varied from $0$ to $1$. In (a), the dashed horizontal line is a guide for the reference to a maximally compact polymer and the white arrow shows the shift in the minima. In (a) and (b), the white arrow shows a marked shift in the behavior of the curves. Horizontal axis labels are share betwen panels (a) and (c) as well as (b) and (d). Error bars  correspond to the standard error of the mean and are generally smaller than the symbol size.}\label{fig:Fig3_SM_Comparison}
\end{figure}

\subsubsection{Chain collapse by preferential mixing of solvent and cosolvent}\label{sec:sc}

Chain collapse from an extended solvated polymer to a dry, solvent-excluded globule is observed in systems where solvent-cosolvent interactions are strongly favorable ($R_\text{sc}$). 
In the isotropic case ($R_\text{sc}^\circ$, blue circles), the collapse coincides with nearly equimolar mixtures at which entropy of mixing is maximized, resulting in a symmetric $R_g$ profile as a function of cosolvent mole fraction (Fig.~\ref{fig:Fig3_SM_Comparison}a); the collapse is likewise evident by the enrichment in monomer-monomer interactions (Fig.~\ref{fig:Fig3_SM_Comparison}b), which is also nearly symmetric. The resulting globule is dry, as indicated by the depletion in excess monomer-cosolvent interactions (Fig.~\ref{fig:Fig3_SM_Comparison}c). 
\rev{The excess number of monomer-cosolvent contacts remain small, indicating behavior close to that expected from ideal mixing and thus limited preferential interaction between cosolvent and polymer.}
Furthermore, the chain is maximally collapsed at $x_\text{c} \approx 0.5$, which aligns with the excess number of solvent-cosolvent interactions (Fig.~\ref{fig:Fig3_SM_Comparison}d).
This dry globule and congruence of polymer collapse with enrichment in solvent-cosolvent interactions are defining characteristics of cononsolvency driven by solvent-cosolvent interactions.

However, there are notable differences 
in the anisotropic case ($R_\text{sc}^\curlywedge$, orange triangles). 
In particular, the maximal chain collapse shifts to lesser cosolvent fractions--in this case, around $x_\text{c}\approx 0.3$ (Fig.~\ref{fig:Fig3_SM_Comparison}a).
This is consistent with the maximum in the number of monomer-monomer contacts (Fig.~\ref{fig:Fig3_SM_Comparison}b), but it is interestingly inconsistent with maximum enrichemnt of solvent-cosolvent interactions, which remains at equimolar concentrations. 
The collapsed state of the polymer remains dry, largely excluding both solvent and cosolvent.
Collectively, these observations highlight that the distinguishing feature between these two systems is the composition at which the polymer is maximally collapsed. 
Whereas both systems should effectively share the same macroscopic phase behavior (at the given temperature), systems with strong anisotropy will be characterized by a more asymmetric $R_g$ profile as a function of cosolvent mole fraction. 

\begin{figure}[htbp]
    \centering
    \includegraphics[keepaspectratio]{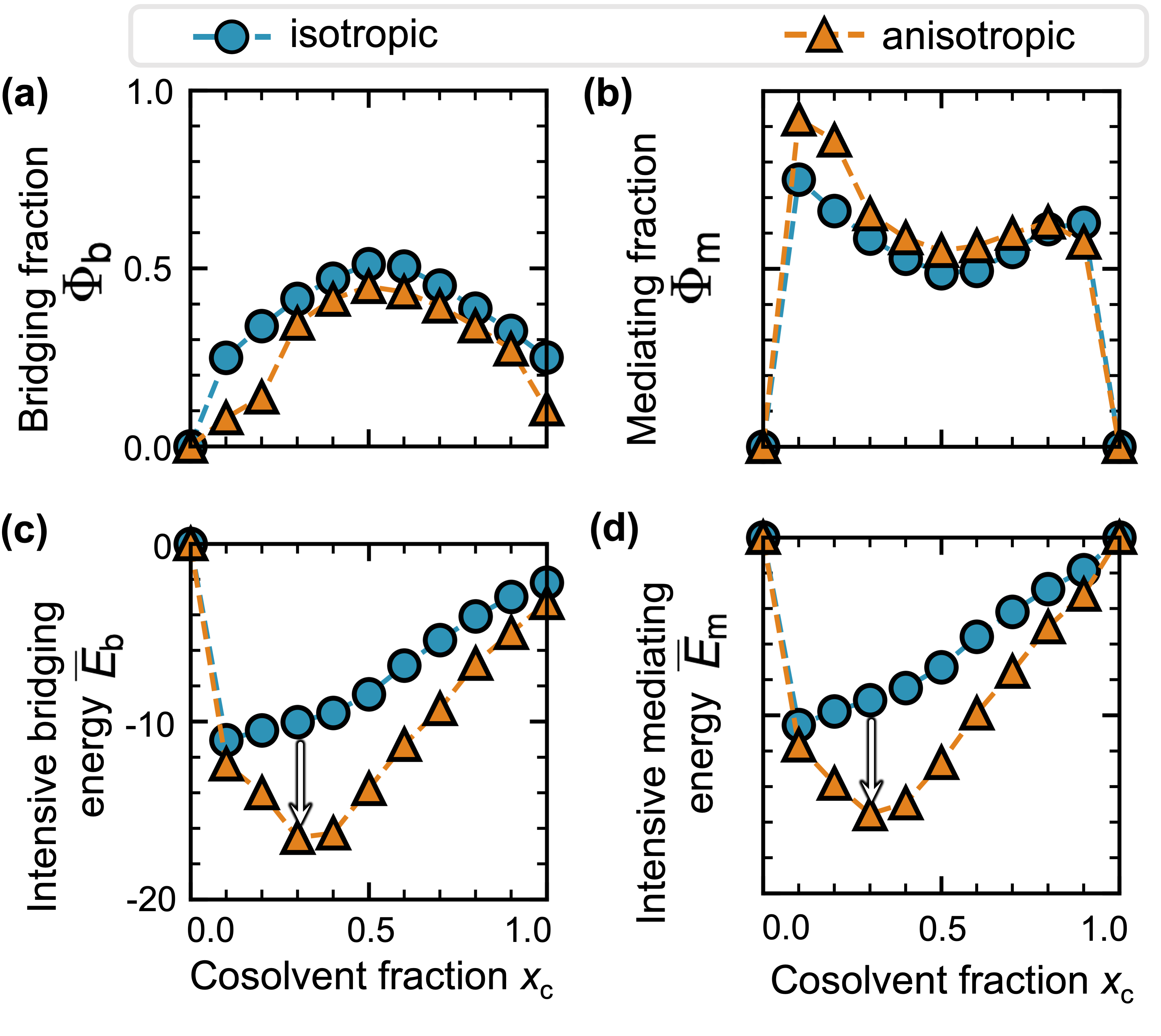}
    \caption{Analysis and comparison of the solvation environment of systems $R_\text{sc}^\circ$ and $R_\text{sc}^\curlywedge$. (a) Fraction of bridging cosolvents $\Phi _\text{b}$, (b) fraction of mediating cosolvents $\Phi _\text{m}$, (c) energetic contribution per cosolvent particle in the solvation shell $\bar{E}_b$, and (d) the energetic contribution per cosolvent particle in the solvation shell $\bar{E}_m$ as cosolvent fraction $x_c$ is varied from $0$ to $1$. Horizontal axis labels are share betwen panels (a) and (c) as well as (b) and (d).  The white arrows highlight a marked difference in the magnitude of intensive energies. Error bars  correspond to the standard error of the mean and are generally smaller than the symbol size.  }\label{fig:Fig4_SM_Cosolvent_Comparison}
\end{figure}

To better understand what drives this asymmetric collapse, we perform a configurational and energetic analysis of the polymer solvation shell. 
In both systems, on trace addition of cosolvent, nearly all cosolvent particles in the solvation shell act as mediating cosolvents, coordinating interactions between monomer and solvent particles, with the remainder acting as bridging cosolvents (Fig.~\ref{fig:Fig4_SM_Cosolvent_Comparison}a,b).
The relative populations of cosolvent particles and how they interact with the polymer is thus not particularly distinctive.
However, the relative energetic contributions from these groups does distinguish the two systems effectively. 
Each bridging and mediating cosolvent in the anisotropic system provides substantially stronger stabilizing interactions to the polymer over certain composition ranges (Fig.~\ref{fig:Fig4_SM_Cosolvent_Comparison}c,d).
\rev{The difference in stabilization between isotropic and anisotropic systems is largest around $x_c \in[0.2, 0.4]$, thereby providing a substantial enthalpic driving force at lower cosolvent fractions and shifting the collapse from $x_\text{c}=0.5$ in the isotropic case to $x_\text{c}=0.3$ in the anisotropic case. }

Ultimately, this elucidates the qualitative difference between single-chain collapse in systems where solvent-cosolvent interactions dominate.
The correlation network formed by bridging and mediating interactions stabilizes the polymer sufficiently to overcome configurational entropy loss, yielding asymmetric collapse. 
By contrast, isotropic systems require maximized solvent mixing to compensate for reduced polymer configurational entropy.

\begin{figure}[htbp]
\centering
\includegraphics[keepaspectratio]{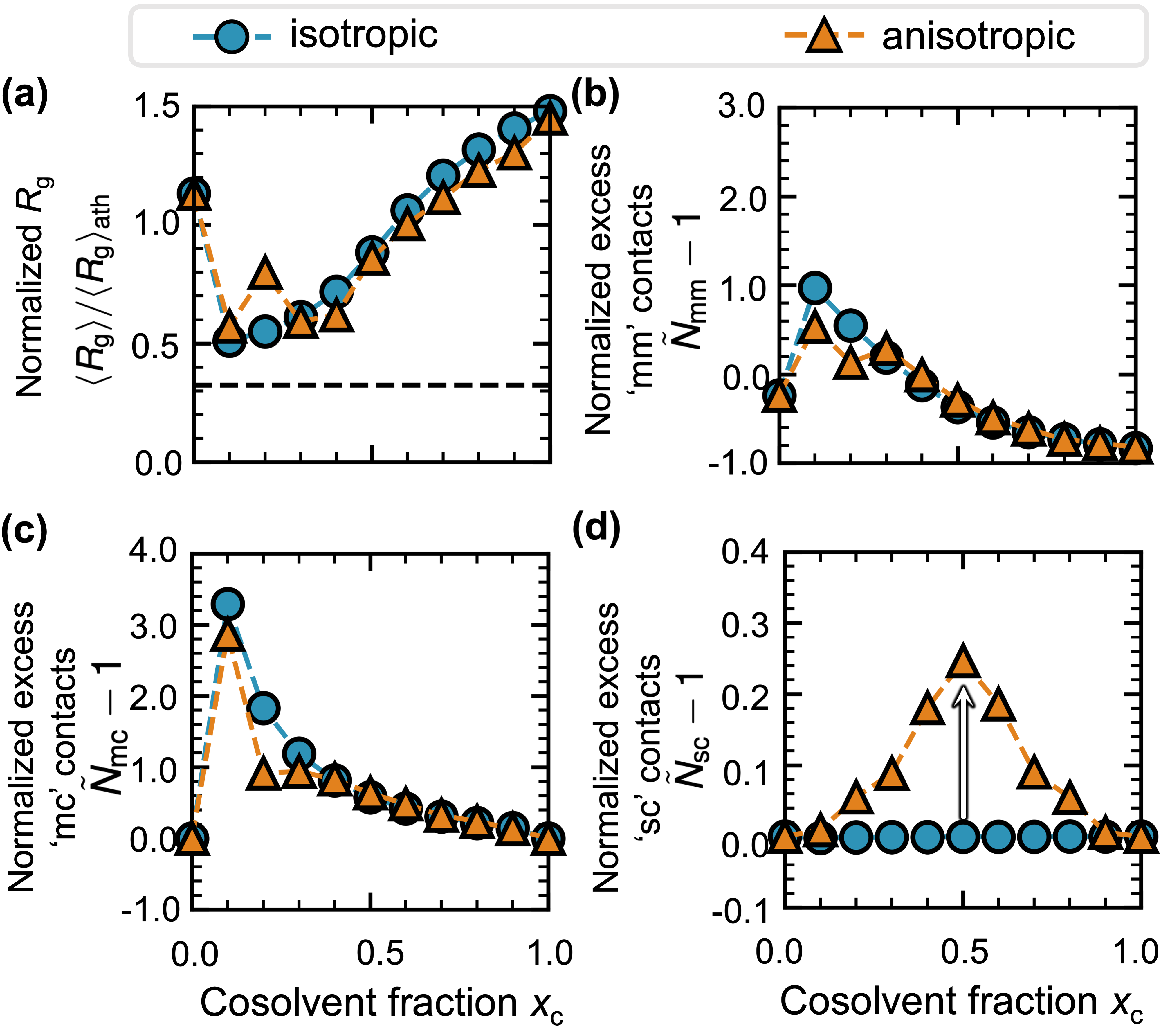}
\caption{A comparison of conformational characteristics and environment of a polymer from $R_\text{pc}^\circ$ (blue circles) and $R_\text{pc}^\curlywedge$ (orange triangles). (a) Normalized single-chain $R_\text{g}$ and normalized excess (b) monomer-monomer, (c) solvent-cosolvent, and (d) monomer-cosolvent interactions as cosolvent fraction $x_c$ is varied from $0$ to $1$. In (a), the dashed horizontal line is a guide for the reference to a maximally compact polymer. Horizontal axis labels are share betwen panels (a) and (c) as well as (b) and (d).  In (c) and (d), the white arrow shows a marked shift in the trendlines. Error bars  correspond to the standard error of the mean and are generally smaller than the symbol size. }\label{fig:Fig5_PreferentialAdsorption}
\end{figure}

\subsubsection{Chain collapse by preferential adsorption of cosolvent}\label{sec:pc}

In a regime dominated by polymer–cosolvent affinity ($R_{\mathrm{pc}}$), most signatures of microscopic cononsolvency are effectively equivalent between systems with equivalent $\chi^\text{FHP}_{ij}$, irrespective of whether such parameters arise from purely isotropic or include anisotropic terms.
For example, both isotropic ($R_{\mathrm{pc}}^\circ$, blue circles) and anisotropic ($R_{\mathrm{pc}}^\curlywedge$ orange triangles) systems display asymmetric cosolvent-induced coil–globule transitions, as monitored by $R_\text{g}$ (Fig.~\ref{fig:Fig5_PreferentialAdsorption}a). 
Notably, this asymmetric profile may be reminiscent of that found for the anisotropic system system in Fig.~\ref{fig:Fig3_SM_Comparison}a.
Nevertheless, by comparing the excess number of monomer-monomer interactions in Fig.~\ref{fig:Fig3_SM_Comparison}b versus those in Fig.~\ref{fig:Fig5_PreferentialAdsorption}b, one can infer that the collapsed chain induced by polymer-cosolvent interactions is relatively wet.
In particular, the globules are found to be cosolvent-laden globule, as evidenced by an excess of cosolvent–monomer interactions  (Fig.~\ref{fig:Fig5_PreferentialAdsorption}c). 
Between isotropic and anisotropic systems, this enrichment is slightly stronger in the latter.
The simulations with anisotropic interactions display more significant solvent-cosolvent interactions at intermediate cosolvent fractions (Fig.~\ref{fig:Fig5_PreferentialAdsorption}d).
However, this enhanced mixing is of little consequence to the polymer conformational behavior and simply arises to due $\Delta_\text{sc}$ being negative.
Because of the strong enrichment in polymer-cosolvent interactions, we mechanistically refer to behavior in this regime as preferential adsorption.

The role of anisotropy for preferential-adsorption driven cononsolvency appears limited to slight differences in solvation motifs. 
On addition of cosolvent, both systems with isotropic and anisotropic interactions possess significant fractions of bridging and mediating cosolvents (Fig.\ref{Fig6:PA_Excess_Ani}a,b). 
By contrast to behavior in the preferential mixing regime, however, here there are notable differences in relative proportions.
For the system with purely isotropic interactions, at low cosolvent fractions, the number of mediating cosolvents exceeds the number of bridging solvents.
However, making interactions with cosolvent orientation-dependent reverses this trend, such that
more cosolvent particles are interacting simultaneously with distal monomers on the polymer chain, rather than being situated between a monomer and solvent.
In conjunction with Fig.~\ref{fig:Fig5_PreferentialAdsorption}c, this implies that the globule in the anisotropic system has more cosolvent embedded or intercalated within its pervaded volume.
Finally, the energetic contribution per cosolvent particle are also notably enhanced when including anisotropic interactions (Fig.\ref{Fig6:PA_Excess_Ani}c,d). 
While this does not seemingly have any clear impact the presence or nature of the coil-globule transition, it may have implications on the properties of the collapsed state.  

Thus, cononsolvency via preferential adsorption exhibits qualitatively similar behavior for both isotropic and anisotropic interactions. In both cases, cosolvent particles percolate the gyration volume, expel solvent, and intercalate between monomer segments to form wet globules. 
\rev{This supports prior observations\cite{R:2014_Mukherji_Polymer} that a polymer chain can collapse without reduction in overall solvent quality. 
    In this case, this arises because the cosolvent intercalates with strong affinity to the polymer chain.}
When interactions are only isotropic, this produces abrupt collapse where cosolvent mediates monomer-solvent interactions; the addition of more cosolvent leads to fewer such interactions, causing gradual re-expansion. 
Anisotropic interactions intensify these physics through enhanced cosolvent intercalation and stronger mediating interactions that persist throughout the transition. 
While the underlying physical drives remain effectively identical, structural and compositional correlations in globules induced by polymer-cosolvent affinity may differ for sufficiently strong anisotropic interactions. 

\begin{figure}[htbp]
\centering
\includegraphics[keepaspectratio]{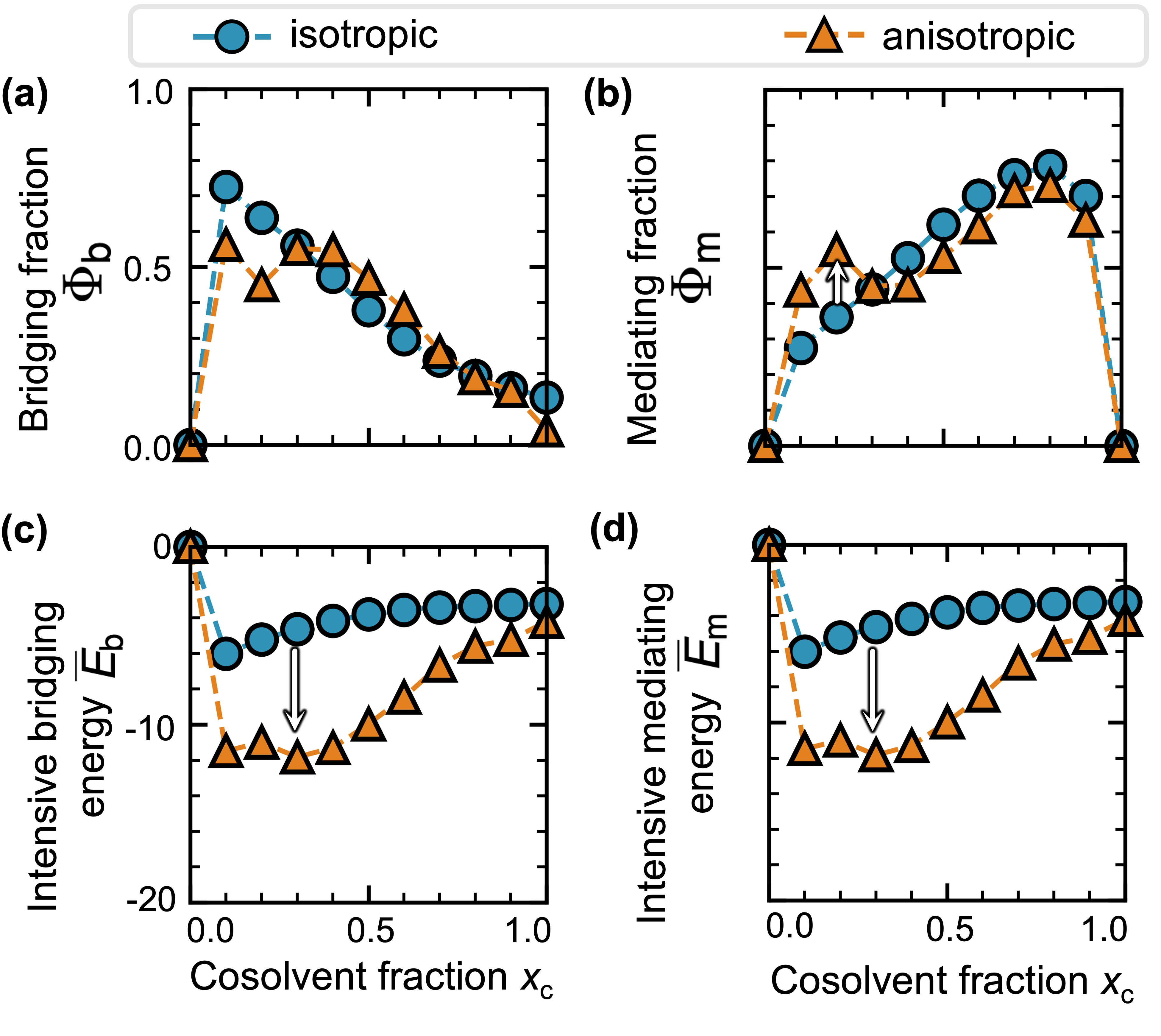}
\caption{
Analysis and comparison of the solvation environment of systems $R_\text{pc}^\circ$ and $R_\text{pc}^\curlywedge$. (a) Fraction of bridging cosolvents $\Phi _\text{b}$, (b) fraction of mediating cosolvents $\Phi _\text{m}$, (c) energetic contribution per cosolvent particle in the solvation shell $\bar{E}_b$, and (d) the energetic contribution per cosolvent particle in the solvation shell $\bar{E}_m$ as cosolvent fraction $x_c$ is varied from $0$ to $1$. Horizontal axis labels are share betwen panels (a) and (c) as well as (b) and (d).  In (c) and (d), the white arrow shows a marked shift in the intensive energy. Error bars  correspond to the standard error of the mean and are generally smaller than the symbol size.}\label{Fig6:PA_Excess_Ani}
\end{figure}

\section{Conclusions}

We investigated how orientational interactions influence cononsolvency phenomena in ternary polymer solutions. 
Using the Flory-Huggins-Potts (FHP) framework enabled systematic comparison of systems with identical effective $\chi$ parameters but different underlying interaction types, including those with orientation-dependent energetic contributions. 
After identifying phase-separation regimes driven by either solvent-cosolvent or polymer-cosolvent affinity, we contrasted systems with the same effective $\chi$ parameters but achieved either purely through isotropic interactions or with inclusion of anisotropic interactions, specifically with cosolvent species.
This approach enabled controlled comparisons between systems with identical mean-field drivers, isolating how anisotropic interactions affect microscopic physics despite equivalent macroscopic phase behavior. Analysis focused on cosolvent-induced coil-globule transitions.
This work complements prior extensive literature on cononsolvency based on FH theory.

Our analysis revealed that \rev{orientation-dependent interactions} have variable importance depending on the dominant interaction type driving cononsolvency. 
For systems driven by strong solvent-cosolvent interactions, orientation-dependent interactions qualitatively altered coil-globule transitions, inducing collapse at lower cosolvent fractions than isotropic systems with equivalent mean-field parameters. 
This enhanced, asymmetric collapse was attributed to stronger cosolvent-mediated interactions, where cosolvent acts as a surfactant favorably interacting with both polymer and solvent.\cite{R:2020_Bharadwaj_cosolvent} 
This asymmetric collapse notably resembled that observed in polymer-cosolvent affinity-driven systems.
However, whereas solvent-cosolvent affinity produced dry, solvent-depleted globules, those produce by polymer-cosolvent affinity were more wet with embedded cosolvent; this distinction offers a pathway to distinguish these regimes at the microscopic level. 
When comparing isotropic and anisotropic systems both dominated by polymer-cosolvent affinity, anisotropic interactions primarily increase bridging-type cosolvent configurations and potentially enhance globule stability, without fundamentally altering the mechanism.

This work highlights several implications and opportunities for future inquiry. Discriminating among cononsolvency mechanisms proves challenging when examining only macroscopic phase behavior or microscopic single-chain conformations in isolation. However, complementary analysis at both scales suggests that preferential mixing, anisotropically-influenced preferential mixing, and preferential adsorption mechanisms can be distinguished. While investigating the role of anisotropy interactions via experiment remains nontrivial, our results suggest that temperature-dependent characterization across compositions may be informative.
While the FHP framework provides physically grounded parameters enabling precise experimental fitting,\cite{R:2024_Dhamankar_Asymmetry} this has so far relied on macroscopic phase behavior. \rev{Connecting FHP parameters to molecular simulations or microscopic experimental observables would clarify when orientation-dependent interactions emerge in real systems. Meanwhile, we propose that studies employing systematic molecular modifications---such as adding electron-withdrawing substituents or stackable molecular moieties to cosolvents to tune hydrogen-bonding and molecular packing---may offer tangible routes to probe anisotropic effects without requiring FHP abstraction.}

\section*{Acknowledgements}
S.D. and M.A.W. acknowledge support from the National Science Foundation under Grant No. 2237470. 
Calculations were performed using resources from Princeton Research Computing at Princeton University, which is a consortium led by the Princeton Institute for Computational Science and Engineering (PICSciE) and Office of Information Technology's Research Computing.

\providecommand*{\mcitethebibliography}{\thebibliography}
\csname @ifundefined\endcsname{endmcitethebibliography}
{\let\endmcitethebibliography\endthebibliography}{}


\begin{mcitethebibliography}{79}
\providecommand*{\natexlab}[1]{#1}
\providecommand*{\mciteSetBstSublistMode}[1]{}
\providecommand*{\mciteSetBstMaxWidthForm}[2]{}
\providecommand*{\mciteBstWouldAddEndPuncttrue}
  {\def\EndOfBibitem{\unskip.}}
\providecommand*{\mciteBstWouldAddEndPunctfalse}
  {\let\EndOfBibitem\relax}
\providecommand*{\mciteSetBstMidEndSepPunct}[3]{}
\providecommand*{\mciteSetBstSublistLabelBeginEnd}[3]{}
\providecommand*{\EndOfBibitem}{}
\mciteSetBstSublistMode{f}
\mciteSetBstMaxWidthForm{subitem}
{(\emph{\alph{mcitesubitemcount}})}
\mciteSetBstSublistLabelBeginEnd{\mcitemaxwidthsubitemform\space}
{\relax}{\relax}

\bibitem[Heskins and Guillet(1968)]{R:1968_Heskins_Solution}
M.~Heskins and J.~E. Guillet, \emph{Journal of Macromolecular Science: Part A - Chemistry}, 1968, \textbf{2}, 1441--1455\relax
\mciteBstWouldAddEndPuncttrue
\mciteSetBstMidEndSepPunct{\mcitedefaultmidpunct}
{\mcitedefaultendpunct}{\mcitedefaultseppunct}\relax
\EndOfBibitem
\bibitem[Bharadwaj \emph{et~al.}(2022)Bharadwaj, Niebuur, Nothdurft, Richtering, van~der Vegt, and Papadakis]{R:2022_Bharadwaj_Cononsolvency}
S.~Bharadwaj, B.-J. Niebuur, K.~Nothdurft, W.~Richtering, N.~F.~A. van~der Vegt and C.~M. Papadakis, \emph{Soft Matter}, 2022, \textbf{18}, 2884--2909\relax
\mciteBstWouldAddEndPuncttrue
\mciteSetBstMidEndSepPunct{\mcitedefaultmidpunct}
{\mcitedefaultendpunct}{\mcitedefaultseppunct}\relax
\EndOfBibitem
\bibitem[Takahashi \emph{et~al.}(2003)Takahashi, Kanaya, Nishida, and Kaji]{R:2003_Takahashi_Effects}
N.~Takahashi, T.~Kanaya, K.~Nishida and K.~Kaji, \emph{Polymer}, 2003, \textbf{44}, 4075--4078\relax
\mciteBstWouldAddEndPuncttrue
\mciteSetBstMidEndSepPunct{\mcitedefaultmidpunct}
{\mcitedefaultendpunct}{\mcitedefaultseppunct}\relax
\EndOfBibitem
\bibitem[Jia \emph{et~al.}(2017)Jia, Muthukumar, Cheng, Han, and Hammouda]{R:2017_Jia_Concentration}
D.~Jia, M.~Muthukumar, H.~Cheng, C.~C. Han and B.~Hammouda, \emph{Macromolecules}, 2017, \textbf{50}, 7291--7298\relax
\mciteBstWouldAddEndPuncttrue
\mciteSetBstMidEndSepPunct{\mcitedefaultmidpunct}
{\mcitedefaultendpunct}{\mcitedefaultseppunct}\relax
\EndOfBibitem
\bibitem[Fernández-Piérola and Horta(1980)]{R:1980_Fernandez-Pierola_Co}
I.~Fernández-Piérola and A.~Horta, \emph{Polymer Bulletin}, 1980, \textbf{3}, 273--278\relax
\mciteBstWouldAddEndPuncttrue
\mciteSetBstMidEndSepPunct{\mcitedefaultmidpunct}
{\mcitedefaultendpunct}{\mcitedefaultseppunct}\relax
\EndOfBibitem
\bibitem[Wang \emph{et~al.}(2012)Wang, Shi, Luo, Chen, and Zhao]{R:2012_Wang_Conformational}
F.~Wang, Y.~Shi, S.~Luo, Y.~Chen and J.~Zhao, \emph{Macromolecules}, 2012, \textbf{45}, 9196--9204\relax
\mciteBstWouldAddEndPuncttrue
\mciteSetBstMidEndSepPunct{\mcitedefaultmidpunct}
{\mcitedefaultendpunct}{\mcitedefaultseppunct}\relax
\EndOfBibitem
\bibitem[Zhang and Wu(2001)]{R:2001_Zhang_Water/Methanol}
G.~Zhang and C.~Wu, \emph{Journal of the American Chemical Society}, 2001, \textbf{123}, 1376--1380\relax
\mciteBstWouldAddEndPuncttrue
\mciteSetBstMidEndSepPunct{\mcitedefaultmidpunct}
{\mcitedefaultendpunct}{\mcitedefaultseppunct}\relax
\EndOfBibitem
\bibitem[Zhao and Kremer(2022)]{R:2022_Zhao_Effects}
Y.~Zhao and K.~Kremer, \emph{Macromolecular Rapid Communications}, 2022, \textbf{43}, 2100907\relax
\mciteBstWouldAddEndPuncttrue
\mciteSetBstMidEndSepPunct{\mcitedefaultmidpunct}
{\mcitedefaultendpunct}{\mcitedefaultseppunct}\relax
\EndOfBibitem
\bibitem[Zhao and Kremer(2021)]{R:2021_Zhao_Proline}
Y.~Zhao and K.~Kremer, \emph{The Journal of Physical Chemistry B}, 2021, \textbf{125}, 9751--9756\relax
\mciteBstWouldAddEndPuncttrue
\mciteSetBstMidEndSepPunct{\mcitedefaultmidpunct}
{\mcitedefaultendpunct}{\mcitedefaultseppunct}\relax
\EndOfBibitem
\bibitem[Hao \emph{et~al.}(2010)Hao, Cheng, Butler, Zhang, and Han]{R:2010_Hao_Origin}
J.~Hao, H.~Cheng, P.~Butler, L.~Zhang and C.~C. Han, \emph{The Journal of Chemical Physics}, 2010, \textbf{132}, 154902\relax
\mciteBstWouldAddEndPuncttrue
\mciteSetBstMidEndSepPunct{\mcitedefaultmidpunct}
{\mcitedefaultendpunct}{\mcitedefaultseppunct}\relax
\EndOfBibitem
\bibitem[Sun and Wu(2010)]{R:2010_Sun_Role}
S.~Sun and P.~Wu, \emph{Macromolecules}, 2010, \textbf{43}, 9501--9510\relax
\mciteBstWouldAddEndPuncttrue
\mciteSetBstMidEndSepPunct{\mcitedefaultmidpunct}
{\mcitedefaultendpunct}{\mcitedefaultseppunct}\relax
\EndOfBibitem
\bibitem[Hore \emph{et~al.}(2013)Hore, Hammouda, Li, and Cheng]{R:2013_Hore_Co}
M.~J.~A. Hore, B.~Hammouda, Y.~Li and H.~Cheng, \emph{Macromolecules}, 2013, \textbf{46}, 7894--7901\relax
\mciteBstWouldAddEndPuncttrue
\mciteSetBstMidEndSepPunct{\mcitedefaultmidpunct}
{\mcitedefaultendpunct}{\mcitedefaultseppunct}\relax
\EndOfBibitem
\bibitem[Jia \emph{et~al.}(2016)Jia, Zuo, Rogers, Cheng, Hammouda, and Han]{R:2016_Jia_Re}
D.~Jia, T.~Zuo, S.~Rogers, H.~Cheng, B.~Hammouda and C.~C. Han, \emph{Macromolecules}, 2016, \textbf{49}, 5152--5159\relax
\mciteBstWouldAddEndPuncttrue
\mciteSetBstMidEndSepPunct{\mcitedefaultmidpunct}
{\mcitedefaultendpunct}{\mcitedefaultseppunct}\relax
\EndOfBibitem
\bibitem[Budkov and Kiselev(2017)]{R:2017_Budkov_Flory}
Y.~A. Budkov and M.~G. Kiselev, \emph{Journal of Physics: Condensed Matter}, 2017, \textbf{30}, 043001\relax
\mciteBstWouldAddEndPuncttrue
\mciteSetBstMidEndSepPunct{\mcitedefaultmidpunct}
{\mcitedefaultendpunct}{\mcitedefaultseppunct}\relax
\EndOfBibitem
\bibitem[Pooch \emph{et~al.}(2019)Pooch, Teltevskij, Karjalainen, Tenhu, and Winnik]{R:2019_Pooch_Poly2}
F.~Pooch, V.~Teltevskij, E.~Karjalainen, H.~Tenhu and F.~M. Winnik, \emph{Macromolecules}, 2019, \textbf{52}, 6361--6368\relax
\mciteBstWouldAddEndPuncttrue
\mciteSetBstMidEndSepPunct{\mcitedefaultmidpunct}
{\mcitedefaultendpunct}{\mcitedefaultseppunct}\relax
\EndOfBibitem
\bibitem[Zuo \emph{et~al.}(2019)Zuo, Ma, Jiao, Han, Xiao, Liang, Hong, Bowron, Soper, Han, and Cheng]{R:2019_Zuo_Water/Cosolvent}
T.~Zuo, C.~Ma, G.~Jiao, Z.~Han, S.~Xiao, H.~Liang, L.~Hong, D.~Bowron, A.~Soper, C.~C. Han and H.~Cheng, \emph{Macromolecules}, 2019, \textbf{52}, 457--464\relax
\mciteBstWouldAddEndPuncttrue
\mciteSetBstMidEndSepPunct{\mcitedefaultmidpunct}
{\mcitedefaultendpunct}{\mcitedefaultseppunct}\relax
\EndOfBibitem
\bibitem[Yong \emph{et~al.}(2019)Yong, Bittrich, Uhlmann, Fery, and Sommer]{R:2019_Yong_Co}
H.~Yong, E.~Bittrich, P.~Uhlmann, A.~Fery and J.-U. Sommer, \emph{Macromolecules}, 2019, \textbf{52}, 6285--6293\relax
\mciteBstWouldAddEndPuncttrue
\mciteSetBstMidEndSepPunct{\mcitedefaultmidpunct}
{\mcitedefaultendpunct}{\mcitedefaultseppunct}\relax
\EndOfBibitem
\bibitem[Zhang \emph{et~al.}(2019)Zhang, Carvalho, Fang, and Serpe]{R:2019_Zhang_Volatile}
Y.~Zhang, W.~S. Carvalho, C.~Fang and M.~J. Serpe, \emph{Sensors and Actuators B: Chemical}, 2019, \textbf{290}, 520--526\relax
\mciteBstWouldAddEndPuncttrue
\mciteSetBstMidEndSepPunct{\mcitedefaultmidpunct}
{\mcitedefaultendpunct}{\mcitedefaultseppunct}\relax
\EndOfBibitem
\bibitem[Nian and Pu(2018)]{R:2018_Nian_Racemic}
S.~Nian and L.~Pu, \emph{The Journal of Organic Chemistry}, 2018, \textbf{84}, 909--913\relax
\mciteBstWouldAddEndPuncttrue
\mciteSetBstMidEndSepPunct{\mcitedefaultmidpunct}
{\mcitedefaultendpunct}{\mcitedefaultseppunct}\relax
\EndOfBibitem
\bibitem[Kleinschmidt \emph{et~al.}(2020)Kleinschmidt, Nothdurft, Anakhov, Meyer, Mork, Gumerov, Potemkin, Richtering, and Pich]{R:2020_Kleinschmidt_Microgel}
D.~Kleinschmidt, K.~Nothdurft, M.~V. Anakhov, A.~A. Meyer, M.~Mork, R.~A. Gumerov, I.~I. Potemkin, W.~Richtering and A.~Pich, \emph{Materials Advances}, 2020, \textbf{1}, 2983--2993\relax
\mciteBstWouldAddEndPuncttrue
\mciteSetBstMidEndSepPunct{\mcitedefaultmidpunct}
{\mcitedefaultendpunct}{\mcitedefaultseppunct}\relax
\EndOfBibitem
\bibitem[Yu \emph{et~al.}(2016)Yu, Cirelli, Kieviet, Kooij, Vancso, and de~Beer]{R:2016_Yu_Tunable}
Y.~Yu, M.~Cirelli, B.~D. Kieviet, E.~S. Kooij, G.~J. Vancso and S.~de~Beer, \emph{Polymer}, 2016, \textbf{102}, 372--378\relax
\mciteBstWouldAddEndPuncttrue
\mciteSetBstMidEndSepPunct{\mcitedefaultmidpunct}
{\mcitedefaultendpunct}{\mcitedefaultseppunct}\relax
\EndOfBibitem
\bibitem[Chen \emph{et~al.}(2014)Chen, Kooij, Sui, Padberg, Hempenius, Schön, and Vancso]{R:2014_Chen_Collapse}
Q.~Chen, E.~S. Kooij, X.~Sui, C.~J. Padberg, M.~A. Hempenius, P.~M. Schön and G.~J. Vancso, \emph{Soft Matter}, 2014, \textbf{10}, 3134\relax
\mciteBstWouldAddEndPuncttrue
\mciteSetBstMidEndSepPunct{\mcitedefaultmidpunct}
{\mcitedefaultendpunct}{\mcitedefaultseppunct}\relax
\EndOfBibitem
\bibitem[Wang \emph{et~al.}(2019)Wang, Huang, Liu, Rehfeldt, Wang, and Zhang]{R:2019_Wang_Multi‐Responsive}
X.~Wang, H.~Huang, H.~Liu, F.~Rehfeldt, X.~Wang and K.~Zhang, \emph{Macromolecular Chemistry and Physics}, 2019, \textbf{220}, \relax
\mciteBstWouldAddEndPuncttrue
\mciteSetBstMidEndSepPunct{\mcitedefaultmidpunct}
{\mcitedefaultendpunct}{\mcitedefaultseppunct}\relax
\EndOfBibitem
\bibitem[Yong \emph{et~al.}(2021)Yong, Molcrette, Sperling, Montel, and Sommer]{R:2021_Yong_Regulating}
H.~Yong, B.~Molcrette, M.~Sperling, F.~Montel and J.-U. Sommer, \emph{Macromolecules}, 2021, \textbf{54}, 4432--4442\relax
\mciteBstWouldAddEndPuncttrue
\mciteSetBstMidEndSepPunct{\mcitedefaultmidpunct}
{\mcitedefaultendpunct}{\mcitedefaultseppunct}\relax
\EndOfBibitem
\bibitem[Mills \emph{et~al.}(2019)Mills, Ding, and Olsen]{R:2019_Mills_Protein}
C.~E. Mills, E.~Ding and B.~Olsen, \emph{Industrial \& Engineering Chemistry Research}, 2019, \textbf{58}, 11698--11709\relax
\mciteBstWouldAddEndPuncttrue
\mciteSetBstMidEndSepPunct{\mcitedefaultmidpunct}
{\mcitedefaultendpunct}{\mcitedefaultseppunct}\relax
\EndOfBibitem
\bibitem[Valente \emph{et~al.}(2005)Valente, Verma, Manning, William~Wilson, and Henry]{R:2005_Valente_Second}
J.~J. Valente, K.~S. Verma, M.~C. Manning, W.~William~Wilson and C.~S. Henry, \emph{Biophysical Journal}, 2005, \textbf{89}, 4211--4218\relax
\mciteBstWouldAddEndPuncttrue
\mciteSetBstMidEndSepPunct{\mcitedefaultmidpunct}
{\mcitedefaultendpunct}{\mcitedefaultseppunct}\relax
\EndOfBibitem
\bibitem[Xiao \emph{et~al.}(2002)Xiao, Gardner, and Sprang]{R:2002_Xiao_Cosolvent}
T.~Xiao, K.~H. Gardner and S.~R. Sprang, \emph{Proceedings of the National Academy of Sciences}, 2002, \textbf{99}, 11151--11156\relax
\mciteBstWouldAddEndPuncttrue
\mciteSetBstMidEndSepPunct{\mcitedefaultmidpunct}
{\mcitedefaultendpunct}{\mcitedefaultseppunct}\relax
\EndOfBibitem
\bibitem[Reddy \emph{et~al.}(2020)Reddy, Muttathukattil, and Mondal]{R:2020_Reddy_Cosolvent}
G.~Reddy, A.~N. Muttathukattil and B.~Mondal, \emph{Current Opinion in Structural Biology}, 2020, \textbf{60}, 101--109\relax
\mciteBstWouldAddEndPuncttrue
\mciteSetBstMidEndSepPunct{\mcitedefaultmidpunct}
{\mcitedefaultendpunct}{\mcitedefaultseppunct}\relax
\EndOfBibitem
\bibitem[Canchi and García(2013)]{R:2013_Canchi_Cosolvent}
D.~R. Canchi and A.~E. García, \emph{Annual Review of Physical Chemistry}, 2013, \textbf{64}, 273--293\relax
\mciteBstWouldAddEndPuncttrue
\mciteSetBstMidEndSepPunct{\mcitedefaultmidpunct}
{\mcitedefaultendpunct}{\mcitedefaultseppunct}\relax
\EndOfBibitem
\bibitem[Gazi \emph{et~al.}(2023)Gazi, Maity, and Jana]{R:2023_Gazi_Conformational}
R.~Gazi, S.~Maity and M.~Jana, \emph{ACS Omega}, 2023, \textbf{8}, 2832--2843\relax
\mciteBstWouldAddEndPuncttrue
\mciteSetBstMidEndSepPunct{\mcitedefaultmidpunct}
{\mcitedefaultendpunct}{\mcitedefaultseppunct}\relax
\EndOfBibitem
\bibitem[Stuart \emph{et~al.}(2010)Stuart, Huck, Genzer, Müller, Ober, Stamm, Sukhorukov, Szleifer, Tsukruk, Urban, Winnik, Zauscher, Luzinov, and Minko]{R:2010_Stuart_Emerging}
M.~A.~C. Stuart, W.~T.~S. Huck, J.~Genzer, M.~Müller, C.~Ober, M.~Stamm, G.~B. Sukhorukov, I.~Szleifer, V.~V. Tsukruk, M.~Urban, F.~Winnik, S.~Zauscher, I.~Luzinov and S.~Minko, \emph{Nature Materials}, 2010, \textbf{9}, 101--113\relax
\mciteBstWouldAddEndPuncttrue
\mciteSetBstMidEndSepPunct{\mcitedefaultmidpunct}
{\mcitedefaultendpunct}{\mcitedefaultseppunct}\relax
\EndOfBibitem
\bibitem[Nettles \emph{et~al.}(2010)Nettles, Chilkoti, and Setton]{R:2010_Nettles_Applications}
D.~L. Nettles, A.~Chilkoti and L.~A. Setton, \emph{Advanced Drug Delivery Reviews}, 2010, \textbf{62}, 1479--1485\relax
\mciteBstWouldAddEndPuncttrue
\mciteSetBstMidEndSepPunct{\mcitedefaultmidpunct}
{\mcitedefaultendpunct}{\mcitedefaultseppunct}\relax
\EndOfBibitem
\bibitem[Rodríguez-Cabello \emph{et~al.}(2016)Rodríguez-Cabello, Arias, Rodrigo, and Girotti]{R:2016_Rodriguez-Cabello_Elastin}
J.~C. Rodríguez-Cabello, F.~J. Arias, M.~A. Rodrigo and A.~Girotti, \emph{Advanced Drug Delivery Reviews}, 2016, \textbf{97}, 85--100\relax
\mciteBstWouldAddEndPuncttrue
\mciteSetBstMidEndSepPunct{\mcitedefaultmidpunct}
{\mcitedefaultendpunct}{\mcitedefaultseppunct}\relax
\EndOfBibitem
\bibitem[Dudowicz \emph{et~al.}(2015)Dudowicz, Freed, and Douglas]{R:2015_Dudowicz_Theory}
J.~Dudowicz, K.~F. Freed and J.~F. Douglas, \emph{The Journal of Chemical Physics}, 2015, \textbf{142}, 214906\relax
\mciteBstWouldAddEndPuncttrue
\mciteSetBstMidEndSepPunct{\mcitedefaultmidpunct}
{\mcitedefaultendpunct}{\mcitedefaultseppunct}\relax
\EndOfBibitem
\bibitem[Wang \emph{et~al.}(2017)Wang, Wang, Liu, Bai, Gong, Ru, and Feng]{R:2017_Wang_Preferential}
J.~Wang, N.~Wang, B.~Liu, J.~Bai, P.~Gong, G.~Ru and J.~Feng, \emph{Physical Chemistry Chemical Physics}, 2017, \textbf{19}, 30097--30106\relax
\mciteBstWouldAddEndPuncttrue
\mciteSetBstMidEndSepPunct{\mcitedefaultmidpunct}
{\mcitedefaultendpunct}{\mcitedefaultseppunct}\relax
\EndOfBibitem
\bibitem[Bharadwaj and van~der Vegt(2019)]{R:2019_Bharadwaj_Does}
S.~Bharadwaj and N.~F.~A. van~der Vegt, \emph{Macromolecules}, 2019, \textbf{52}, 4131--4138\relax
\mciteBstWouldAddEndPuncttrue
\mciteSetBstMidEndSepPunct{\mcitedefaultmidpunct}
{\mcitedefaultendpunct}{\mcitedefaultseppunct}\relax
\EndOfBibitem
\bibitem[Schild \emph{et~al.}(1991)Schild, Muthukumar, and Tirrell]{R:1991_Schild_Cononsolvency}
H.~G. Schild, M.~Muthukumar and D.~A. Tirrell, \emph{Macromolecules}, 1991, \textbf{24}, 948--952\relax
\mciteBstWouldAddEndPuncttrue
\mciteSetBstMidEndSepPunct{\mcitedefaultmidpunct}
{\mcitedefaultendpunct}{\mcitedefaultseppunct}\relax
\EndOfBibitem
\bibitem[Mukherji \emph{et~al.}(2015)Mukherji, Marques, Stuehn, and Kremer]{R:2015_Mukherji_Co}
D.~Mukherji, C.~M. Marques, T.~Stuehn and K.~Kremer, \emph{The Journal of Chemical Physics}, 2015, \textbf{142}, 114903\relax
\mciteBstWouldAddEndPuncttrue
\mciteSetBstMidEndSepPunct{\mcitedefaultmidpunct}
{\mcitedefaultendpunct}{\mcitedefaultseppunct}\relax
\EndOfBibitem
\bibitem[Tanaka(1985)]{R:1985_Tanaka_polymer}
F.~Tanaka, \emph{The Journal of Chemical Physics}, 1985, \textbf{82}, 2466--2471\relax
\mciteBstWouldAddEndPuncttrue
\mciteSetBstMidEndSepPunct{\mcitedefaultmidpunct}
{\mcitedefaultendpunct}{\mcitedefaultseppunct}\relax
\EndOfBibitem
\bibitem[Tanaka \emph{et~al.}(1982)Tanaka, Nishio, Sun, and Ueno-Nishio]{R:1982_Tanaka_Collapse}
T.~Tanaka, I.~Nishio, S.-T. Sun and S.~Ueno-Nishio, \emph{Science}, 1982, \textbf{218}, 467--469\relax
\mciteBstWouldAddEndPuncttrue
\mciteSetBstMidEndSepPunct{\mcitedefaultmidpunct}
{\mcitedefaultendpunct}{\mcitedefaultseppunct}\relax
\EndOfBibitem
\bibitem[Tanaka \emph{et~al.}(2008)Tanaka, Koga, and Winnik]{R:2008_Tanaka_Temperature}
F.~Tanaka, T.~Koga and F.~M. Winnik, \emph{Physical Review Letters}, 2008, \textbf{101}, 028302\relax
\mciteBstWouldAddEndPuncttrue
\mciteSetBstMidEndSepPunct{\mcitedefaultmidpunct}
{\mcitedefaultendpunct}{\mcitedefaultseppunct}\relax
\EndOfBibitem
\bibitem[Tanaka \emph{et~al.}(2011)Tanaka, Koga, Kojima, Xue, and Winnik]{R:2011_Tanaka_Preferential}
F.~Tanaka, T.~Koga, H.~Kojima, N.~Xue and F.~M. Winnik, \emph{Macromolecules}, 2011, \textbf{44}, 2978--2989\relax
\mciteBstWouldAddEndPuncttrue
\mciteSetBstMidEndSepPunct{\mcitedefaultmidpunct}
{\mcitedefaultendpunct}{\mcitedefaultseppunct}\relax
\EndOfBibitem
\bibitem[Dalgicdir and van~der Vegt(2019)]{R:2019_Dalgicdir_Improved}
C.~Dalgicdir and N.~F.~A. van~der Vegt, \emph{The Journal of Physical Chemistry B}, 2019, \textbf{123}, 3875--3883\relax
\mciteBstWouldAddEndPuncttrue
\mciteSetBstMidEndSepPunct{\mcitedefaultmidpunct}
{\mcitedefaultendpunct}{\mcitedefaultseppunct}\relax
\EndOfBibitem
\bibitem[Pérez-Ramírez \emph{et~al.}(2019)Pérez-Ramírez, Haro-Pérez, Vázquez-Contreras, Klapp, Bautista-Carbajal, and Odriozola]{R:2019_Perez-Ramirez_P}
H.~A. Pérez-Ramírez, C.~Haro-Pérez, E.~Vázquez-Contreras, J.~Klapp, G.~Bautista-Carbajal and G.~Odriozola, \emph{Physical Chemistry Chemical Physics}, 2019, \textbf{21}, 5106--5116\relax
\mciteBstWouldAddEndPuncttrue
\mciteSetBstMidEndSepPunct{\mcitedefaultmidpunct}
{\mcitedefaultendpunct}{\mcitedefaultseppunct}\relax
\EndOfBibitem
\bibitem[Budkov and Kolesnikov(2018)]{R:2018_Budkov_Models}
Y.~A. Budkov and A.~L. Kolesnikov, \emph{Polymer Science, Series C}, 2018, \textbf{60}, 148--159\relax
\mciteBstWouldAddEndPuncttrue
\mciteSetBstMidEndSepPunct{\mcitedefaultmidpunct}
{\mcitedefaultendpunct}{\mcitedefaultseppunct}\relax
\EndOfBibitem
\bibitem[Budkov and Kolesnikov(2017)]{R:2017_Budkov_Statistical}
Y.~A. Budkov and A.~L. Kolesnikov, \emph{Soft Matter}, 2017, \textbf{13}, 8362--8367\relax
\mciteBstWouldAddEndPuncttrue
\mciteSetBstMidEndSepPunct{\mcitedefaultmidpunct}
{\mcitedefaultendpunct}{\mcitedefaultseppunct}\relax
\EndOfBibitem
\bibitem[Bharadwaj \emph{et~al.}(2020)Bharadwaj, Nayar, Dalgicdir, and van~der Vegt]{R:2020_Bharadwaj_cosolvent}
S.~Bharadwaj, D.~Nayar, C.~Dalgicdir and N.~F.~A. van~der Vegt, \emph{Communications Chemistry}, 2020, \textbf{3}, 1--7\relax
\mciteBstWouldAddEndPuncttrue
\mciteSetBstMidEndSepPunct{\mcitedefaultmidpunct}
{\mcitedefaultendpunct}{\mcitedefaultseppunct}\relax
\EndOfBibitem
\bibitem[Pica and Graziano(2016)]{R:2016_Pica_alternative}
A.~Pica and G.~Graziano, \emph{Physical Chemistry Chemical Physics}, 2016, \textbf{18}, 25601--25608\relax
\mciteBstWouldAddEndPuncttrue
\mciteSetBstMidEndSepPunct{\mcitedefaultmidpunct}
{\mcitedefaultendpunct}{\mcitedefaultseppunct}\relax
\EndOfBibitem
\bibitem[Dalgicdir \emph{et~al.}(2017)Dalgicdir, Rodríguez-Ropero, and van~der Vegt]{R:2017_Dalgicdir_Computational}
C.~Dalgicdir, F.~Rodríguez-Ropero and N.~F.~A. van~der Vegt, \emph{The Journal of Physical Chemistry B}, 2017, \textbf{121}, 7741--7748\relax
\mciteBstWouldAddEndPuncttrue
\mciteSetBstMidEndSepPunct{\mcitedefaultmidpunct}
{\mcitedefaultendpunct}{\mcitedefaultseppunct}\relax
\EndOfBibitem
\bibitem[Dudowicz \emph{et~al.}(2015)Dudowicz, Freed, and Douglas]{R:2015_Dudowicz_Communication:}
J.~Dudowicz, K.~F. Freed and J.~F. Douglas, \emph{The Journal of Chemical Physics}, 2015, \textbf{143}, 131101\relax
\mciteBstWouldAddEndPuncttrue
\mciteSetBstMidEndSepPunct{\mcitedefaultmidpunct}
{\mcitedefaultendpunct}{\mcitedefaultseppunct}\relax
\EndOfBibitem
\bibitem[Zhang(2024)]{R:2024_Zhang_Phase}
P.~Zhang, \emph{Macromolecules}, 2024, \textbf{57}, 4298--4311\relax
\mciteBstWouldAddEndPuncttrue
\mciteSetBstMidEndSepPunct{\mcitedefaultmidpunct}
{\mcitedefaultendpunct}{\mcitedefaultseppunct}\relax
\EndOfBibitem
\bibitem[Zhang \emph{et~al.}(2020)Zhang, Zong, and Meng]{R:2020_Zhang_unified}
X.~Zhang, J.~Zong and D.~Meng, \emph{Soft Matter}, 2020, \textbf{16}, 7789--7796\relax
\mciteBstWouldAddEndPuncttrue
\mciteSetBstMidEndSepPunct{\mcitedefaultmidpunct}
{\mcitedefaultendpunct}{\mcitedefaultseppunct}\relax
\EndOfBibitem
\bibitem[Oliver \emph{et~al.}(2025)Oliver, Jacobs, and Webb]{R:2025_Oliver_When}
W.~W. Oliver, W.~M. Jacobs and M.~A. Webb, \emph{The Journal of Physical Chemistry B}, 2025, \textbf{129}, 9551--9565\relax
\mciteBstWouldAddEndPuncttrue
\mciteSetBstMidEndSepPunct{\mcitedefaultmidpunct}
{\mcitedefaultendpunct}{\mcitedefaultseppunct}\relax
\EndOfBibitem
\bibitem[Liu \emph{et~al.}(2024)Liu, Duan, and Wang]{R:2024_Liu_Variational}
L.~Liu, C.~Duan and R.~Wang, \emph{Macromolecules}, 2024, \textbf{57}, 10694--10703\relax
\mciteBstWouldAddEndPuncttrue
\mciteSetBstMidEndSepPunct{\mcitedefaultmidpunct}
{\mcitedefaultendpunct}{\mcitedefaultseppunct}\relax
\EndOfBibitem
\bibitem[Bharadwaj \emph{et~al.}(2021)Bharadwaj, Nayar, Dalgicdir, and van~der Vegt]{R:2021_Bharadwaj_interplay}
S.~Bharadwaj, D.~Nayar, C.~Dalgicdir and N.~F.~A. van~der Vegt, \emph{The Journal of Chemical Physics}, 2021, \textbf{154}, 134903\relax
\mciteBstWouldAddEndPuncttrue
\mciteSetBstMidEndSepPunct{\mcitedefaultmidpunct}
{\mcitedefaultendpunct}{\mcitedefaultseppunct}\relax
\EndOfBibitem
\bibitem[Mukherji \emph{et~al.}(2014)Mukherji, Marques, and Kremer]{R:2014_Mukherji_Polymer}
D.~Mukherji, C.~M. Marques and K.~Kremer, \emph{Nature Communications}, 2014, \textbf{5}, 1--6\relax
\mciteBstWouldAddEndPuncttrue
\mciteSetBstMidEndSepPunct{\mcitedefaultmidpunct}
{\mcitedefaultendpunct}{\mcitedefaultseppunct}\relax
\EndOfBibitem
\bibitem[Sommer(2017)]{R:2017_Sommer_AdsorptionAttraction}
J.-U. Sommer, \emph{Macromolecules}, 2017, \textbf{50}, 2219--2228\relax
\mciteBstWouldAddEndPuncttrue
\mciteSetBstMidEndSepPunct{\mcitedefaultmidpunct}
{\mcitedefaultendpunct}{\mcitedefaultseppunct}\relax
\EndOfBibitem
\bibitem[Sommer(2018)]{R:2018_Sommer_Gluonic}
J.-U. Sommer, \emph{Macromolecules}, 2018, \textbf{51}, 3066--3074\relax
\mciteBstWouldAddEndPuncttrue
\mciteSetBstMidEndSepPunct{\mcitedefaultmidpunct}
{\mcitedefaultendpunct}{\mcitedefaultseppunct}\relax
\EndOfBibitem
\bibitem[Mukherji and Kremer(2013)]{R:2013_Mukherji_Coil–Globule–Coil}
D.~Mukherji and K.~Kremer, \emph{Macromolecules}, 2013, \textbf{46}, 9158--9163\relax
\mciteBstWouldAddEndPuncttrue
\mciteSetBstMidEndSepPunct{\mcitedefaultmidpunct}
{\mcitedefaultendpunct}{\mcitedefaultseppunct}\relax
\EndOfBibitem
\bibitem[Mukherji \emph{et~al.}(2016)Mukherji, Wagner, Watson, Winzen, de~Oliveira, Marques, and Kremer]{R:2016_Mukherji_Relating}
D.~Mukherji, M.~Wagner, M.~D. Watson, S.~Winzen, T.~E. de~Oliveira, C.~M. Marques and K.~Kremer, \emph{Soft Matter}, 2016, \textbf{12}, 7995--8003\relax
\mciteBstWouldAddEndPuncttrue
\mciteSetBstMidEndSepPunct{\mcitedefaultmidpunct}
{\mcitedefaultendpunct}{\mcitedefaultseppunct}\relax
\EndOfBibitem
\bibitem[Pica and Graziano(2017)]{R:2017_Pica_Comment}
A.~Pica and G.~Graziano, \emph{Soft Matter}, 2017, \textbf{13}, 7698--7700\relax
\mciteBstWouldAddEndPuncttrue
\mciteSetBstMidEndSepPunct{\mcitedefaultmidpunct}
{\mcitedefaultendpunct}{\mcitedefaultseppunct}\relax
\EndOfBibitem
\bibitem[van~der Vegt and Rodríguez-Ropero(2017)]{R:2017_Vegt_Comment}
N.~F.~A. van~der Vegt and F.~Rodríguez-Ropero, \emph{Soft Matter}, 2017, \textbf{13}, 2289--2291\relax
\mciteBstWouldAddEndPuncttrue
\mciteSetBstMidEndSepPunct{\mcitedefaultmidpunct}
{\mcitedefaultendpunct}{\mcitedefaultseppunct}\relax
\EndOfBibitem
\bibitem[Mukherji \emph{et~al.}(2017)Mukherji, Wagner, Watson, Winzen, de~Oliveira, Marques, and Kremer]{R:2017_Mukherji_Reply}
D.~Mukherji, M.~Wagner, M.~D. Watson, S.~Winzen, T.~E. de~Oliveira, C.~M. Marques and K.~Kremer, \emph{Soft Matter}, 2017, \textbf{13}, 7701--7703\relax
\mciteBstWouldAddEndPuncttrue
\mciteSetBstMidEndSepPunct{\mcitedefaultmidpunct}
{\mcitedefaultendpunct}{\mcitedefaultseppunct}\relax
\EndOfBibitem
\bibitem[Zhang(2025)]{R:2025_Zhang_Phase}
P.~Zhang, \emph{Macromolecules}, 2025, \textbf{58}, 2472--2483\relax
\mciteBstWouldAddEndPuncttrue
\mciteSetBstMidEndSepPunct{\mcitedefaultmidpunct}
{\mcitedefaultendpunct}{\mcitedefaultseppunct}\relax
\EndOfBibitem
\bibitem[Zhang \emph{et~al.}(2022)Zhang, Wang, and Wang]{R:2022_Zhang_Conformation}
P.~Zhang, Z.~Wang and Z.-G. Wang, \emph{Macromolecules}, 2022, \textbf{56}, 153--165\relax
\mciteBstWouldAddEndPuncttrue
\mciteSetBstMidEndSepPunct{\mcitedefaultmidpunct}
{\mcitedefaultendpunct}{\mcitedefaultseppunct}\relax
\EndOfBibitem
\bibitem[Marcato \emph{et~al.}(2024)Marcato, Giacometti, Maritan, and Rosa]{R:2024_Marcato_Theory}
D.~Marcato, A.~Giacometti, A.~Maritan and A.~Rosa, \emph{Physical Review Materials}, 2024, \textbf{8}, 125601\relax
\mciteBstWouldAddEndPuncttrue
\mciteSetBstMidEndSepPunct{\mcitedefaultmidpunct}
{\mcitedefaultendpunct}{\mcitedefaultseppunct}\relax
\EndOfBibitem
\bibitem[Li \emph{et~al.}(2025)Li, Wang, Wang, Yin, Jiang, Zhang, and Li]{R:2025_Li_novel}
X.~Li, Z.~Wang, Z.~Wang, Y.~Yin, R.~Jiang, P.~Zhang and B.~Li, \emph{Soft Matter}, 2025, \textbf{21}, 4858--4868\relax
\mciteBstWouldAddEndPuncttrue
\mciteSetBstMidEndSepPunct{\mcitedefaultmidpunct}
{\mcitedefaultendpunct}{\mcitedefaultseppunct}\relax
\EndOfBibitem
\bibitem[Scherzinger \emph{et~al.}(2014)Scherzinger, Schwarz, Bardow, Leonhard, and Richtering]{R:2014_Scherzinger_Cononsolvency}
C.~Scherzinger, A.~Schwarz, A.~Bardow, K.~Leonhard and W.~Richtering, \emph{Current Opinion in Colloid Interface Science}, 2014, \textbf{19}, 84--94\relax
\mciteBstWouldAddEndPuncttrue
\mciteSetBstMidEndSepPunct{\mcitedefaultmidpunct}
{\mcitedefaultendpunct}{\mcitedefaultseppunct}\relax
\EndOfBibitem
\bibitem[Yong and Sommer(2022)]{R:2022_Yong_Cononsolvency}
H.~Yong and J.-U. Sommer, \emph{Macromolecules}, 2022, \textbf{55}, 11034--11050\relax
\mciteBstWouldAddEndPuncttrue
\mciteSetBstMidEndSepPunct{\mcitedefaultmidpunct}
{\mcitedefaultendpunct}{\mcitedefaultseppunct}\relax
\EndOfBibitem
\bibitem[Dhamankar and Webb(2024)]{R:2024_Dhamankar_Asymmetry}
S.~Dhamankar and M.~A. Webb, \emph{ACS Macro Letters}, 2024, \textbf{13}, 818--825\relax
\mciteBstWouldAddEndPuncttrue
\mciteSetBstMidEndSepPunct{\mcitedefaultmidpunct}
{\mcitedefaultendpunct}{\mcitedefaultseppunct}\relax
\EndOfBibitem
\bibitem[van~der Schoot(2022)]{R:2022_Schoot_Molecular}
P.~van~der Schoot, \emph{Molecular Theory of Nematic (and Other) Liquid Crystals: An Introduction}, Springer International Publishing, 2022\relax
\mciteBstWouldAddEndPuncttrue
\mciteSetBstMidEndSepPunct{\mcitedefaultmidpunct}
{\mcitedefaultendpunct}{\mcitedefaultseppunct}\relax
\EndOfBibitem
\bibitem[Dixon \emph{et~al.}(1994)Dixon, Dobbs, and Valentini]{R:1994_Dixon_Amide}
D.~A. Dixon, K.~D. Dobbs and J.~J. Valentini, \emph{The Journal of Physical Chemistry}, 1994, \textbf{98}, 13435--13439\relax
\mciteBstWouldAddEndPuncttrue
\mciteSetBstMidEndSepPunct{\mcitedefaultmidpunct}
{\mcitedefaultendpunct}{\mcitedefaultseppunct}\relax
\EndOfBibitem
\bibitem[Huber \emph{et~al.}(2014)Huber, Margreiter, Fuchs, von Grafenstein, Tautermann, Liedl, and Fox]{R:2014_Huber_Heteroaromatic}
R.~G. Huber, M.~A. Margreiter, J.~E. Fuchs, S.~von Grafenstein, C.~S. Tautermann, K.~R. Liedl and T.~Fox, \emph{Journal of Chemical Information and Modeling}, 2014, \textbf{54}, 1371--1379\relax
\mciteBstWouldAddEndPuncttrue
\mciteSetBstMidEndSepPunct{\mcitedefaultmidpunct}
{\mcitedefaultendpunct}{\mcitedefaultseppunct}\relax
\EndOfBibitem
\bibitem[Molčanov and Kojić-Prodić(2019)]{R:2019_Molcanov_Towards}
K.~Molčanov and B.~Kojić-Prodić, \emph{IUCrJ}, 2019, \textbf{6}, 156--166\relax
\mciteBstWouldAddEndPuncttrue
\mciteSetBstMidEndSepPunct{\mcitedefaultmidpunct}
{\mcitedefaultendpunct}{\mcitedefaultseppunct}\relax
\EndOfBibitem
\bibitem[Finneran \emph{et~al.}(2015)Finneran, Carroll, Allodi, and Blake]{R:2015_Finneran_Hydrogen}
I.~A. Finneran, P.~B. Carroll, M.~A. Allodi and G.~A. Blake, \emph{Physical Chemistry Chemical Physics}, 2015, \textbf{17}, 24210--24214\relax
\mciteBstWouldAddEndPuncttrue
\mciteSetBstMidEndSepPunct{\mcitedefaultmidpunct}
{\mcitedefaultendpunct}{\mcitedefaultseppunct}\relax
\EndOfBibitem
\bibitem[Swathi \emph{et~al.}(2022)Swathi, Abdulkareem, Kartha, and Madhurima]{R:2022_Swathi_Hydrogen}
P.~V. Swathi, U.~Abdulkareem, T.~R. Kartha and V.~Madhurima, \emph{ChemistrySelect}, 2022, \textbf{7}, e202200413\relax
\mciteBstWouldAddEndPuncttrue
\mciteSetBstMidEndSepPunct{\mcitedefaultmidpunct}
{\mcitedefaultendpunct}{\mcitedefaultseppunct}\relax
\EndOfBibitem
\bibitem[Scott(1949)]{R:1949_Scott_Thermodynamics}
R.~L. Scott, \emph{The Journal of Chemical Physics}, 1949, \textbf{17}, 268--279\relax
\mciteBstWouldAddEndPuncttrue
\mciteSetBstMidEndSepPunct{\mcitedefaultmidpunct}
{\mcitedefaultendpunct}{\mcitedefaultseppunct}\relax
\EndOfBibitem
\bibitem[Tompa(1949)]{R:1949_Tompa_Critical}
H.~Tompa, \emph{The Journal of Chemical Physics}, 1949, \textbf{17}, 1006--1006\relax
\mciteBstWouldAddEndPuncttrue
\mciteSetBstMidEndSepPunct{\mcitedefaultmidpunct}
{\mcitedefaultendpunct}{\mcitedefaultseppunct}\relax
\EndOfBibitem
\bibitem[Dhamankar \emph{et~al.}(2025)Dhamankar, Jiang, and Webb]{R:2025_Dhamankar_Accelerating}
S.~Dhamankar, S.~Jiang and M.~A. Webb, \emph{Molecular Systems Design \& Engineering}, 2025, \textbf{10}, 89--101\relax
\mciteBstWouldAddEndPuncttrue
\mciteSetBstMidEndSepPunct{\mcitedefaultmidpunct}
{\mcitedefaultendpunct}{\mcitedefaultseppunct}\relax
\EndOfBibitem
\end{mcitethebibliography}
\end{document}


\title{Supplementary Information \\
for\\
Role of interaction anisotropy in polymer cononsolvency: insights from the Flory-Huggins-Potts framework}

\author{Satyen Dhamankar$^1$ and Michael A. Webb${}^1{}^*$\\
\\
{\small $^1$Department of Chemical and Biological Engineering, Princeton University, Princeton, NJ 08544}\\
{\small $^*$Corresponding Author:  mawebb@princeton.edu}}

\date{}
\maketitle

\tableofcontents

\newpage

\section{Monte Carlo simulation details}\label{sec:SImontecarlo}

Simulations are conducted on a $60\times 60\times 60$ simple cubic lattice with periodic boundary conditions. The lattice is populated by solvent particles and a single polymer chain of length $N_\text{m}=72$.
We employ a coordination number $z=26$, incorporating interactions with nearest, next-nearest, and next-next-nearest neighbors. Particle orientations are restricted to directions pointing toward these 26 neighboring sites.
Monte Carlo (MC) moves for solvent particles include exchanges of orientation between neighboring solvent sites and collective orientation perturbations.
For the polymer chain, MC updates involve end-rotation, reptation, orientation changes, and chain regrowth accompanied by orientation perturbations.
The full set of MC moves is summarized in Table {\ref{tab_SI:mcmoves}}, with each move type selected at equal probability.
Chain regrowth moves are performed using a Rosenbluth sampling method to improve sampling efficiency.
Each simulation runs for $10^8$ MC moves, with the second half used for data analysis. Reported averages are based on 30 independent simulations, and all error bars denote the standard error of the mean.

\newpage

\begin{table}[H]
    \centering
    \caption{Summary of Monte Carlo moves employed in the simulations. Moves either modify the position and/or orientation of selected particles on the lattice.}
    \resizebox{\textwidth}{!}{
    \normalsize
    \begin{tabular}{|c|c|c|}
         \hline
         & Move type & Move description  \\
         \hline
         \hline 
         1 & \makecell[l]{End rotation} & \makecell[l]{The polymer end segments, $m_1$ and $m_M$, are randomly selected and attempt to exchange \\ with a neighboring solvent site adjacent to $m_2$ or $m_{M-1}$, respectively.} \\
         \hline
         2 & \makecell[l]{Reptation \\ (`slithering snake')} & \makecell[l]{An end monomer ($m_1$ or $m_M$) is chosen uniformly, and the polymer is shifted in a direction. \\ The move is accepted if the chain advances into its other end’s former location \\ or displaces a solvent particle.} \\
         \hline
         3 & \makecell[l]{Chain regrowth \\ with orientation \\ adjustment} & \makecell[l]{A monomer index $i$ between $2$ and $M-1$ is picked uniformly. \\If $|M-i| < |i-1|$, monomers $m_{i+1}, m_{i+2}, \ldots, m_M$ are repositioned and reoriented;\\ if $|M-i| > |i-1|$, then monomers $m_1, m_2, \ldots, m_{i-1}$ are modified instead.\\ If $|M-i| = |i-1|$, one of the two possible segments is randomly selected.} \\
         \hline
         4 & \makecell[l]{Chain regrowth \\ without orientation \\ adjustment} & \makecell[l]{A monomer index $i$ between $2$ and $M-1$ is selected uniformly.\\If $|M-i| < |i-1|$, monomers $m_{i+1}, m_{i+2},\ldots,m_M$ are repositioned only;\\ if $|M-i| > |i-1|$, monomers $m_1, m_2, \ldots, m_{i-1}$ are repositioned only;\\ if $|M-i| = |i-1|$, either set is chosen at random.} \\
         \hline
         5 & \makecell[l]{Polymer orientation\\ perturbation} & \makecell[l]{A random subset of polymer monomers has its orientations randomly perturbed.} \\
         \hline 
         6 & \makecell[l]{Solvent orientation\\ perturbation} & \makecell[l]{Among the $(L/2)^3$ solvent particles, $n$ are randomly selected and their orientations are perturbed.} \\
         \hline
         7 & \makecell[l]{Solvent exchange} & \makecell[l]{Two solvent particles are chosen randomly and their positions are swapped.} \\
         \hline
         8 & \makecell[l]{Solvent exchange \\ with orientation \\ perturbation} & \makecell[l]{Two solvent particles are randomly picked; their positions are exchanged \\ and their orientations are randomly perturbed.} \\ 
         \hline 
         9 & \makecell[l]{Solvation shell \\ perturbation} & \makecell[l]{Solvent particles directly contacting the polymer have their orientations perturbed.} \\
         \hline 
         10 & \makecell[l]{Solvation shell \\ exchange} & \makecell[l]{A solvent particle near the polymer and another particle elsewhere in the system \\ are selected and their positions are swapped.} \\ 
         \hline 
         11 & \makecell[l]{Solvation shell \\ exchange with \\ orientation perturbation} & \makecell[l]{A solvent particle near the polymer and a randomly chosen solvent particle elsewhere \\ swap locations and both have their orientations randomly perturbed.} \\
         \hline 
    \end{tabular}}
    \label{tab_SI:mcmoves}
\end{table}

\newpage
\section{Representative Phase Diagrams}

\begin{figure*}[h]
\centering
\includegraphics{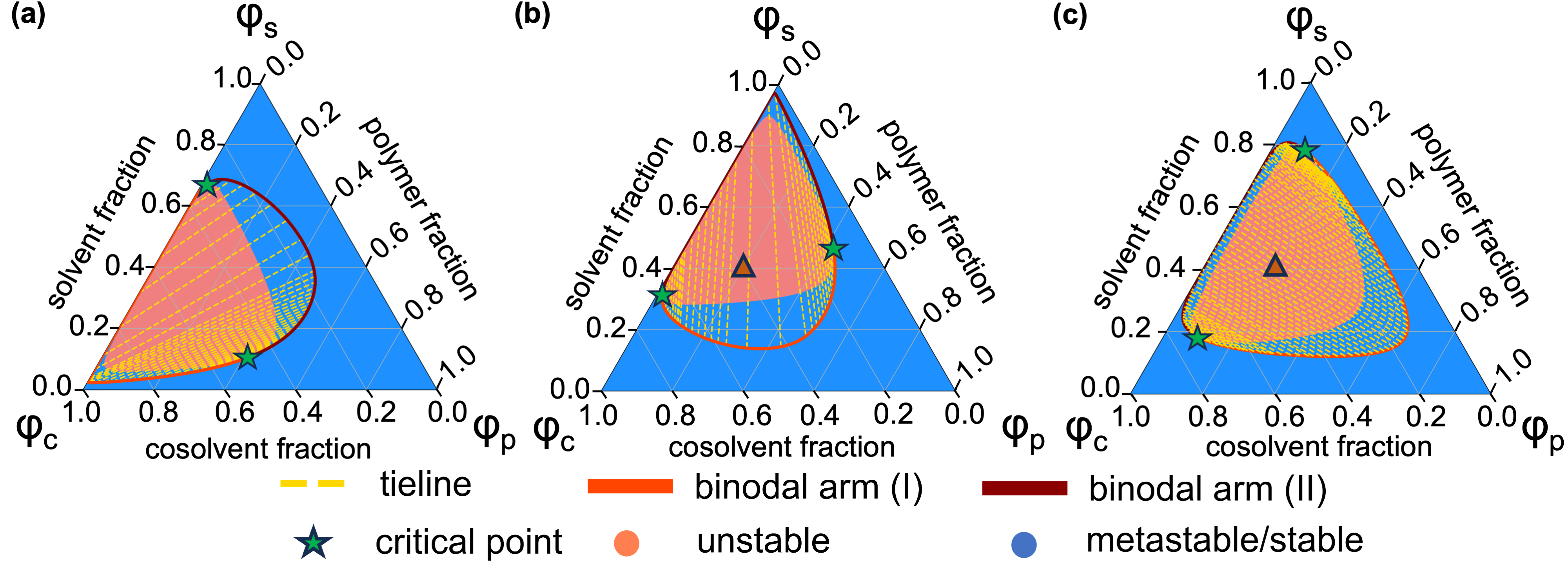}
\caption{Ternary phase diagrams for representative systems that exhibit phase separation in the FHP framework. The specific parameter sets $\boldsymbol{X}^\text{FHP} = (\chi^\text{FHP}_\text{ps}, \chi^\text{FHP}_\text{pc}, \chi^\text{FHP}_\text{sc})$ are drawn from the regions (a) $\mathcal{R}_\text{ps}$ with $\boldsymbol{X}^\text{FHP} =(-10, -1, 0)$, (b) $\mathcal{R}_\text{pc}$ with $\boldsymbol{X}^\text{FHP} = (-1, -10, 0)$, and (c) $\mathcal{R}_\text{sc}$ with $\boldsymbol{X}^\text{FHP} = (-1, -2, -10)$.  The positions of these parameter sets are shown as blue squares in Fig. 1. A representative set of tie lines (yellow, dashed lines) are shown. The tie lines connect points of specific compositions on two distinct binodal arms (I and II) where  the chemical potentials of the constituent species are equal, resulting in equilibrium phase coexistence. The green stars indicate critical points. The peach coloring highlights regions of instability for a homogeneous mixture (i.e., spinodal regions). The composition marked with ochre triangles are the points where an eigenmode analysis of the Hessian was performed. }\label{FigS2:Binodals}
\end{figure*}

\section{Eigenmode Analysis of the Hessian}
To further characterize the instability predicted by the sign of the determinant of the Hessian, we analyzed the unstable eigenmodes of the free energy surface of the ternary system. 
We computed spinodal boundaries by solving for composition using $\chi$ values obtained from the stability criterion 
\begin{align}
    \det \mathbf{H} = 0
\end{align}
where $\mathbf{H}$ is the Hessian matrix of the Helmholtz free energy. 

The resulting binodals indicate binary coexistence compositions under mutual miscibility. Specifically, in three distinct regimes:
\begin{enumerate}
    \item $\chi_\text{sc}$-dominated regimes (strong solvent-cosolvent affinity), $\mathcal{R}_\text{sc}$.
    \item $\chi_\text{pc}$-dominated regimes (strong polymer-cosolvent affinity), $\mathcal{R}_\text{pc}$.
    \item $\chi_\text{ps}$-dominated regimes (strong polymer-solvent affinity), $\mathcal{R}_\text{ps}$.
\end{enumerate}
We only analyze $\mathcal{R}_\text{sc}$ and $\mathcal{R}_\text{pc}$ as the regimes $\mathcal{R}_\text{pc},\mathcal{R}_\text{ps}$ are symmetric under species exchange. To confirm the physical nature of these instabilities, we performed an eigenmode analysis of the Hessian to determine the dominant fluctuation channels.

For the Hessian $\mathbf{H}$, we compute the eigenvectors and eigenvalues $,\mathbf{v}_i,\lambda_i)$ and if all $\lambda _i>0$, the system is stable, and if one $\lambda _i < 0$, the system is unstable along the corresponding eigenvector $\mathbf{v}_i$.

For a ternary incompressible system,
\begin{align*}
    \phi _\text{p} + \phi _\text{s} + \phi _\text{c} &= 1 \\ 
    \delta \phi _\text{p} + \delta \phi _\text{s} + \delta \phi _\text{c} &= 0 
\end{align*}

Given an eigenvector $\mathbf{v}=(\delta \phi_\text{p},\delta \phi_\text{s})$, we reconstruct $\delta \phi_\text{c}=-\delta \phi_\text{p}-\delta \phi _\text{s}$ to obtain the volume fluctuation vector 
\begin{align*}
    \mathbf{v} = (\delta \phi_\text{p},\delta \phi _\text{s}, \delta \phi _\text{c})
\end{align*}
We classify unstable modes by projecting $\mathbf{v}$ onto three canonical fluctuation channels: 
\begin{itemize}
    \item $\mathbf{m}_1=(1,-0.5,-0.5)$: polymer fluctuates opposite to the combined solvent and cosolvent composition.
    \item $\mathbf{m}_2 = (-0.5, 1, -0.5)$: polymer and cosolvent fluctuate together, opposite to the solvent.
    \item $\mathbf{m}_3 = (-0.5, -0.5, 1)$: polymer and solvent fluctuate together, opposite to the cosolvent. 
\end{itemize}
The dominant channel is identified by 
\begin{align}
    i^* = \arg \max _i  \frac{|\mathbf{v} \cdot \mathbf{m}_i |}{\|\mathbf{m}_i\|}
\end{align}

We evaluated a representative composition inside the spinodal $(\phi_\text{p},\phi _\text{c})=(0.2,0.4)$ using $v_\text{p}=72, v_\text{s}=v_\text{c}=1$. 

\begin{table}[h!]
\centering
\caption{Eigenmode analysis in the $\mathcal{R}_\text{pc}$ regime (\(\chi_{\text{pc}}\)-dominated collapse). 
}
\begin{tabular}{cccccc}
\hline
\textbf{Mode} & \(\lambda\) & \([\delta\phi_\text{p},\delta\phi_\text{s}]\) & \(\delta\phi_\text{c}\) & \textbf{Channel} & \textbf{Stability} \\
\hline
Unstable & \(-0.6866\)& \([+0.4432,\,-0.8963]\)& \(+0.4531\)& \((\text{p}+\text{c})|\text{s}\) & Unstable \\
Stable   & \(28.256\)& \([-0.8963,\,-0.4432]\)& \(+1.3396\)& --- & Stable \\
\hline
\end{tabular}
\label{tab:rpc_regime}
\end{table}
The composition analyzed is \((\phi_\text{p},\phi_\text{s}) = (0.2,0.4)\). 
The unstable mode corresponds to the \((p+c)|s\) channel, indicating that the polymer and cosolvent cluster together, expelling solvent.

\begin{table}[h!]
\centering
\caption{Eigenmode analysis in the $\mathcal{R}_\text{sc}$ regime (\(\chi_{\text{sc}}\)-dominated collapse). 
}
\begin{tabular}{cccccc}
\hline
\textbf{Mode} & \(\lambda\) & \([\delta\phi_p,\delta\phi_s]\) & \(\delta\phi_c\) & \textbf{Channel} & \textbf{Stability} \\
\hline
Unstable & \(-0.5607\)& \([-0.8842,\,+0.4670]\)& \(+0.4172\)& \(\text{p}|(\text{s}+\text{c})\) & Unstable \\
Stable   & \(32.130\)& \([+0.4670,\,+0.8842]\)& \(-1.3512\)& --- & Stable \\
\hline
\end{tabular}
\label{tab:rsc_regime}
\end{table}
The composition analyzed is \((\phi_p,\phi_s) = (0.2,0.4)\). 
The unstable mode corresponds to the \(\text{p}|(\text{s}+\text{c})\) channel, indicating that polymer-rich regions segregate from the mixed solvent phase.

\newpage
\section{Robustness and Convergence Checks}

\begin{figure}[H]
    \centering
     \includegraphics[keepaspectratio]{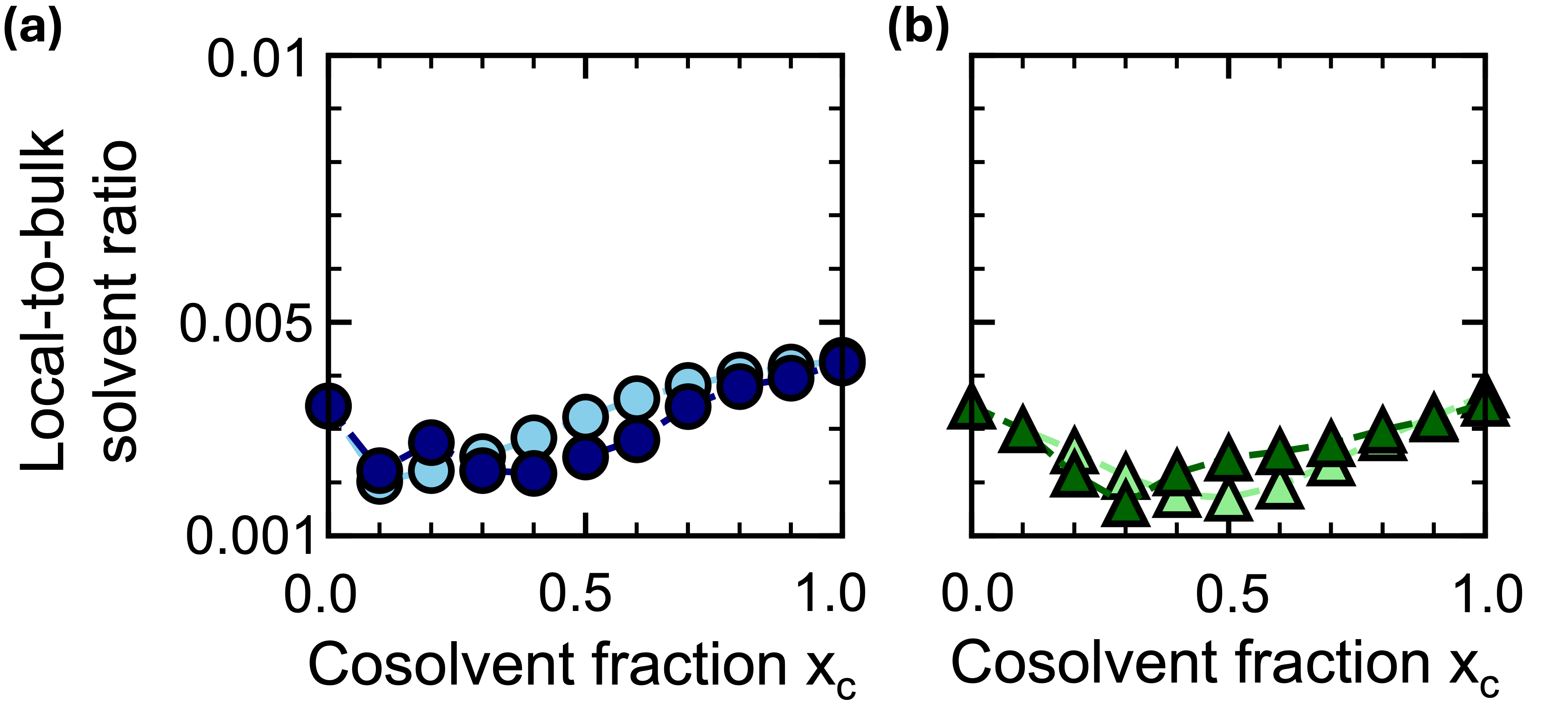}
     \caption{Fraction of total number of solvent and cosolvent particles in contact with the polymer to the number of particles in the bulk. (a) The light and dark blue circles represent computations from isotropic and anistropic simulations where solvent mixing is emphasized. (b) The light and dark green triangles represent computations from isotropic and anistropic simulations where preferential adsorption of cosolvent is emphasized.}
\end{figure}

\begin{figure}[H]
    \centering
    \includegraphics[keepaspectratio]{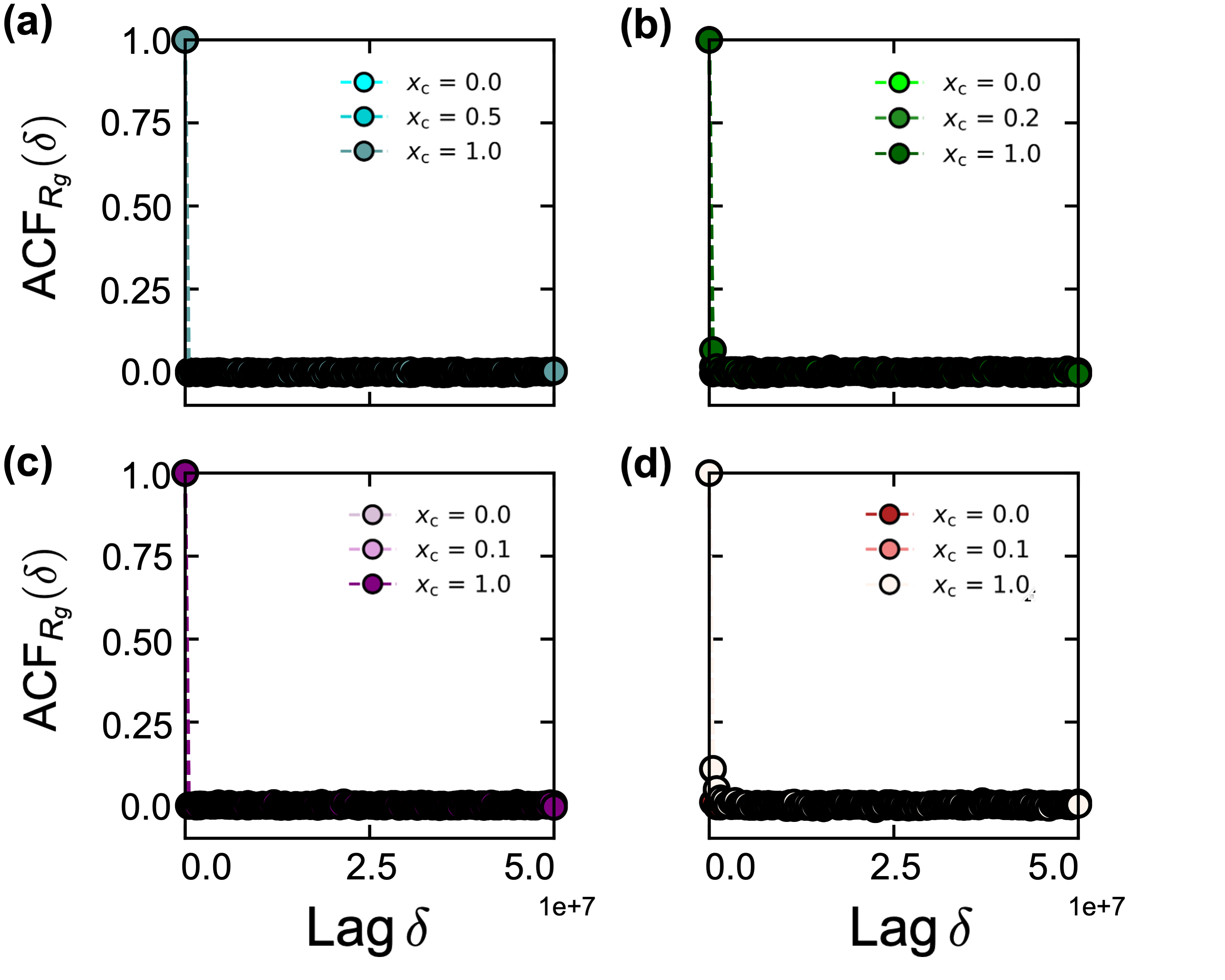}
    \caption{Autocorrelation of the radius of gyration during the final $5\cdot10^7$ Monte Carlo moves. Autocorrelation function of (a) $\mathcal{R}_\text{sc}^\circ$ at $x_\text{c} = 0.0, 0.5$, $1.0$, (b) $\mathcal{R}_\text{sc}^\curlywedge$ at $x_\text{c} = 0.0 0.2$, and $1.0$, (c) $\mathcal{R}_\text{pc}^\circ$ at $x_\text{c} = 0.0, 0.1$, and $1.0$ and (d) $\mathcal{R}_\text{pc}^\curlywedge$ at $x_c=0.0,0.1,1.0$. Results are from thirty independent simulations. Error bars (standard error) are smaller than the symbols.
}
\end{figure}

\begin{figure}[H]
    \centering
    \includegraphics[keepaspectratio]{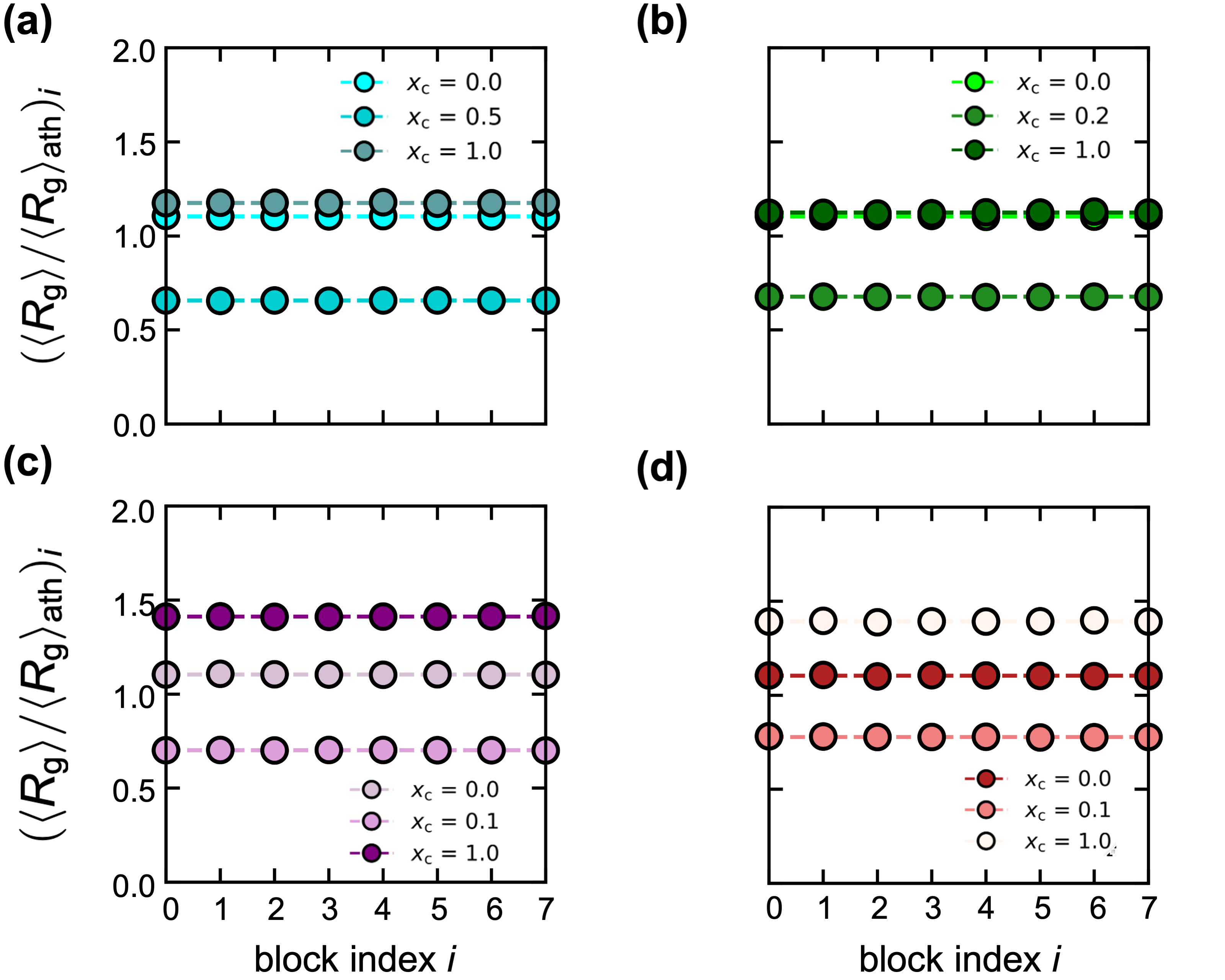}
    \caption{Block averages of the normalized radius of gyration calculated over the final $5\cdot10^7$ Monte Carlo moves (eight equal‑length blocks). Block averages of normalized radius of gyration of (a) $\mathcal{R}_\text{sc}^\circ$ at $x_\text{c} = 0.0, 0.5$, $1.0$, (b) $\mathcal{R}_\text{sc}^\curlywedge$ at $x_\text{c} = 0.0 0.2$, and $1.0$, (c) $\mathcal{R}_\text{pc}^\circ$ at $x_\text{c} = 0.0, 0.1$, and $1.0$ and (d) $\mathcal{R}_\text{pc}^\curlywedge$ at $x_c=0.0,0.1,1.0$. Results are from thirty independent simulations. Error bars (standard error) are smaller than the symbols.
    }
\end{figure}

\newpage
\section{Details of Microscopic Models}

\begin{table*}[htbp]
  \centering
  \caption{Parameters used in anisotropic Monte Carlo simulations of FHP models. Parameters for label type $\mathcal{R}_{ij}^\curlywedge$ with $\Delta _{ij}=-0.2,-0.8$. }\label{tab:sim_parameters}
  \setlength{\tabcolsep}{12pt}

  \renewcommand{\arraystretch}{1.6}
  
  \scriptsize
  \begin{tabular*}{\textwidth}{|c|ccc||cccc|}
    \cline{1-8}
    \textbf{Label} & $\chi^\mathrm{FHP}_{\text{ps}}$ & $\chi^\mathrm{FHP}_{\text{pc}}$ & $\chi^\mathrm{FHP}_{\text{sc}}$ 
    & $\epsilon^{\nparallel}_{\text{mc}}$ & $\epsilon^{\nparallel}_{\text{sc}}$ 
    & $\Delta_{\text{mc}}$ & $\Delta_{\text{sc}}$ \\
    \cline{1-8}
    $\mathcal{R}_{\text{sc}}^\curlywedge$
      & -1  & -2     & -10 
      & -0.5254 & -0.3588
      & -0.2   & -0.2  \\
    \cline{1-8}
    $\mathcal{R}_{\text{sc}}^\curlywedge$
      & -1        & -2      & -10 
      & -0.2426 & -0.0759
      & -0.8   & -0.8\\
    \cline{1-8}\cline{1-8}\cline{1-8}
    $\mathcal{R}_{\text{pc}}^\curlywedge$
      & -1  & -10    & 0   
      & -0.8588 & 0.0579
      & -0.2   & -0.2 \\
    \cline{1-8}
    $\mathcal{R}_{\text{pc}}^\curlywedge$
      & -1  & -10    & 0 
      & -0.5759 & 0.3407
      & -0.8   & -0.8\\
      \cline{1-8}
  \end{tabular*}
\end{table*}

\newpage
\section{Additional Results for Other Parameter Combinations}
\begin{figure}[H]
    \centering
    \includegraphics[keepaspectratio]{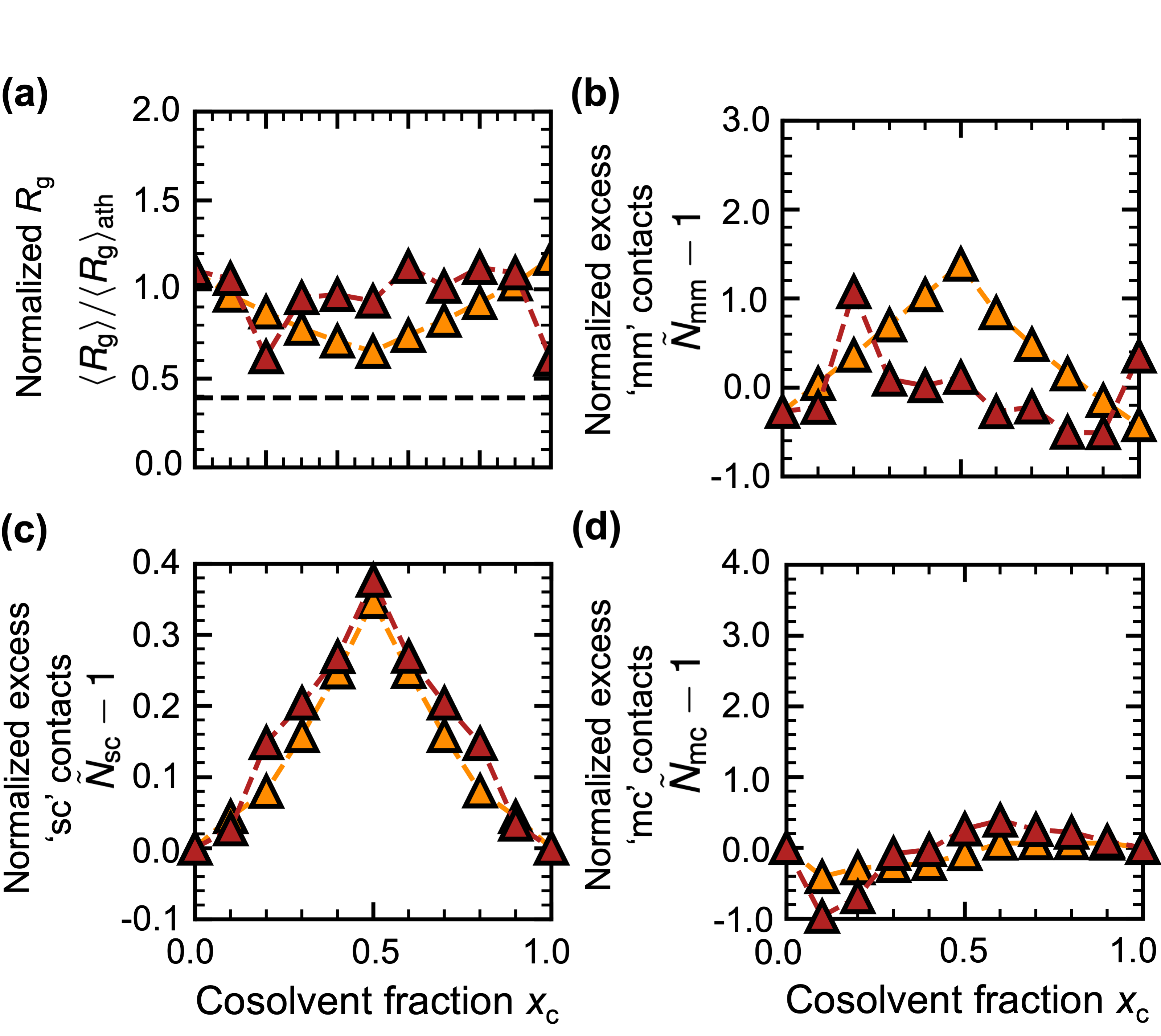}
    \caption{A comparison of conformational characteristics and environment of a polymer from $\mathcal{R}_\text{sc}^\curlywedge$ with $\Delta=-0.2$ (orange) and $\Delta=-0.8$ (dark red). (a) Normalized single-chain $R_\text{g}$ and normalized excess (b) monomer-monomer, (c) solvent-cosolvent, and (d) monomer-cosolvent interactions as cosolvent fraction $x_c$ is varied from $0$ to $1$. In (a), the dashed horizontal line is a guide for the reference to a maximally compact polymer. In all panels the horizontal axis denotes the cosolvent mole fraction \(x_{\mathrm{c}}\). Error bars  correspond to the standard error of the mean and are generally smaller than the symbol size. These simulations are run for a chain with $N_\text{m}=32$.}
\end{figure}

\begin{figure}[H]
    \centering
    \includegraphics[keepaspectratio]{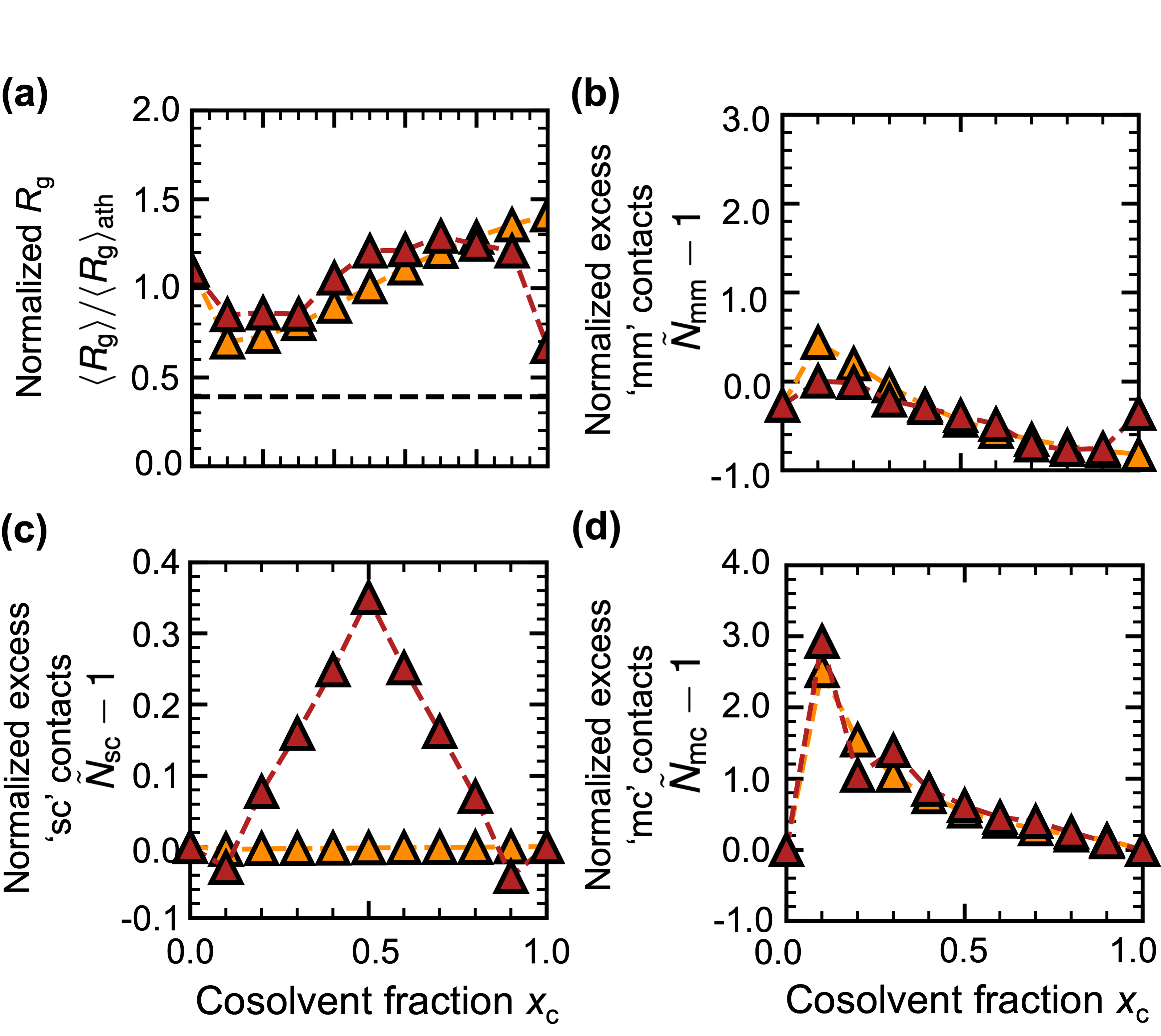}
    \caption{Analysis and comparison of the solvation environment of systems $\mathcal{R}_\text{sc}^\circ$ with $\Delta=-0.2$ (orange) and $\Delta=-0.8$ (dark red). (a) Fraction of bridging cosolvents $\Phi _b$, (b) fraction of mediating cosolvents $\Phi _m$, (c) energetic contribution per cosolvent particle in the solvation shell $\bar{E}_b$, and (d) the energetic contribution per cosolvent particle in the solvation shell $\bar{E}_m$ as cosolvent fraction $x_c$ is varied from $0$ to $1$. In all panels the horizontal axis denotes the cosolvent mole fraction \(x_{\mathrm{c}}\). Error bars  correspond to the standard error of the mean and are generally smaller than the symbol size. These simulations are run for a chain with $N_\text{m}=32$.}
\end{figure}

\begin{figure}[H]
    \centering
    \includegraphics[keepaspectratio]{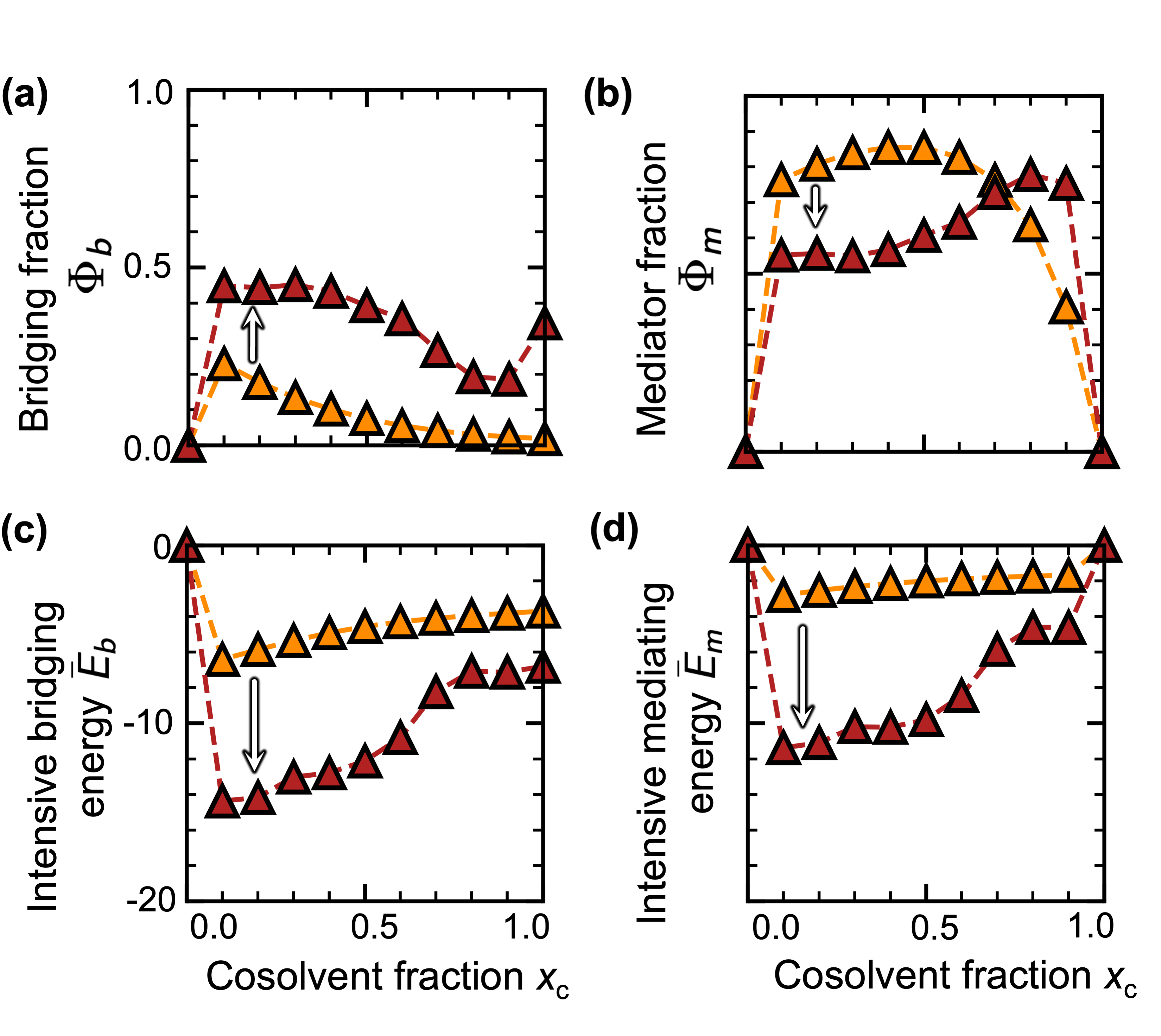}
    \caption{A comparison of conformational characteristics and environment of a polymer from $\mathcal{R}_\text{pc}^\curlywedge$ with $\Delta=-0.2$ (orange) and $\Delta=-0.8$ (dark red). (a) Normalized single-chain $R_\text{g}$ and normalized excess (b) monomer-monomer, (c) solvent-cosolvent, and (d) monomer-cosolvent interactions as cosolvent fraction $x_c$ is varied from $0$ to $1$. In (a), the dashed horizontal line is a guide for the reference to a maximally compact polymer. In all panels the horizontal axis denotes the cosolvent mole fraction \(x_{\mathrm{c}}\). Error bars  correspond to the standard error of the mean and are generally smaller than the symbol size. These simulations are run for a chain with $N_\text{m}=32$.}
\end{figure}

\begin{figure}[H]
    \centering
    \includegraphics[keepaspectratio]{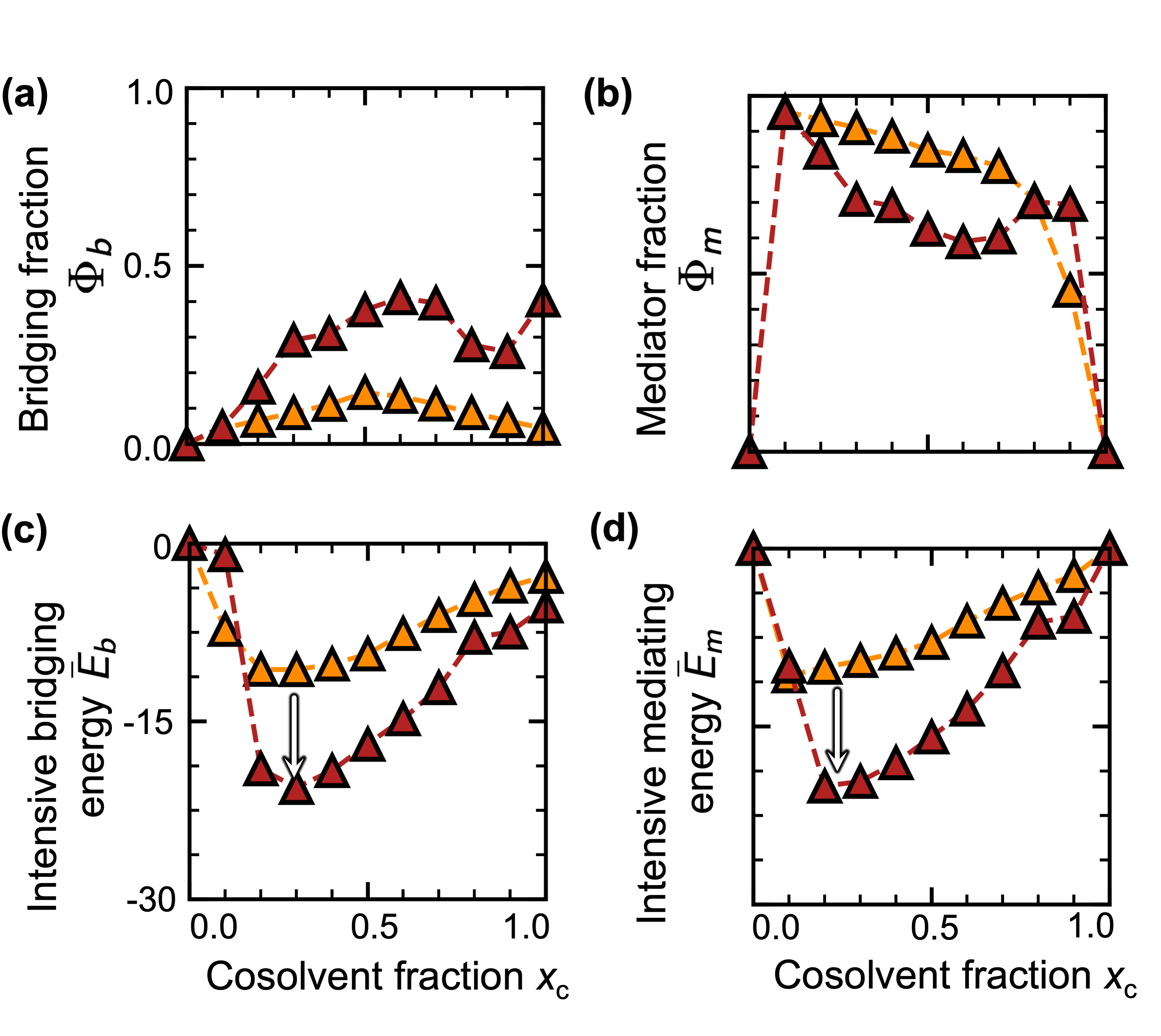}
    \caption{Analysis and comparison of the solvation environment of systems $\mathcal{R}_\text{pc}^\circ$ with $\Delta=-0.2$ (orange) and $\Delta=-0.8$ (dark red). (a) Fraction of bridging cosolvents $\Phi _b$, (b) fraction of mediating cosolvents $\Phi _m$, (c) energetic contribution per cosolvent particle in the solvation shell $\bar{E}_b$, and (d) the energetic contribution per cosolvent particle in the solvation shell $\bar{E}_m$ as cosolvent fraction $x_c$ is varied from $0$ to $1$. In all panels the horizontal axis denotes the cosolvent mole fraction \(x_{\mathrm{c}}\). Error bars  correspond to the standard error of the mean and are generally smaller than the symbol size. These simulations are run for a chain with $N_\text{m}=32$.}
\end{figure}

\begin{figure}[H]
    \centering
    \includegraphics[keepaspectratio]{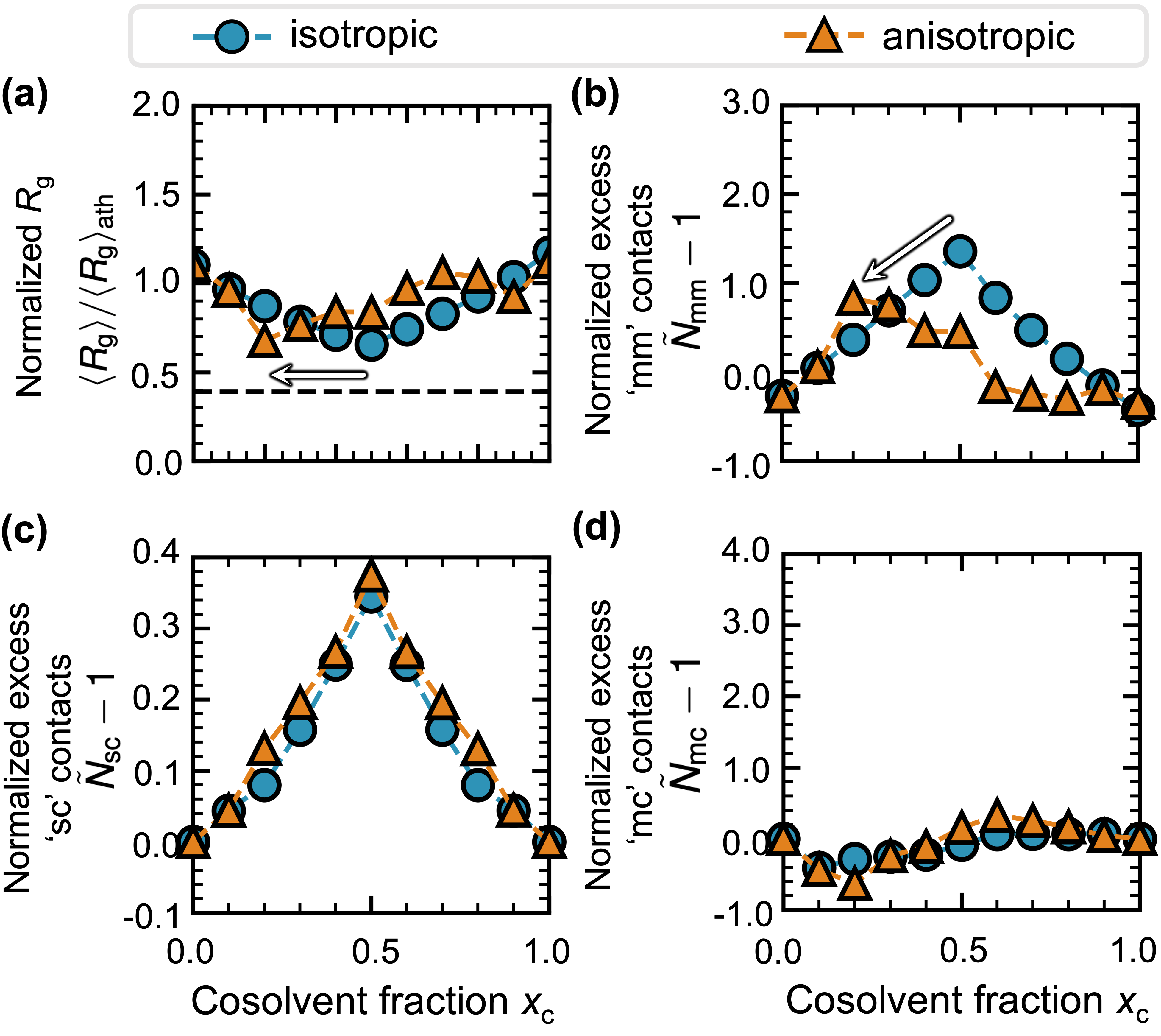}
    \caption{A comparison of conformational characteristics and environment of a polymer from $R_\text{sc}^\circ$ (blue circles) and $R_\text{sc}^\curlywedge$ (orange triangles) for $N_\text{m}=32$. (a) Normalized single-chain $R_\text{g}$ and normalized excess (b) monomer-monomer, (c) monomer-cosolvent and (d) solvent-cosolvent interactions as cosolvent fraction $x_c$ is varied from $0$ to $1$. In (a), the dashed horizontal line is a guide for the reference to a maximally compact polymer and the white arrow shows the shift in the minima. In (a) and (b), the white arrow shows a marked shift in the behavior of the curves. Horizontal axis labels are share betwen panels (a) and (c) as well as (b) and (d). Error bars  correspond to the standard error of the mean and are generally smaller than the symbol size.}
\end{figure}

\begin{figure}[H]
    \centering
    \includegraphics[keepaspectratio]{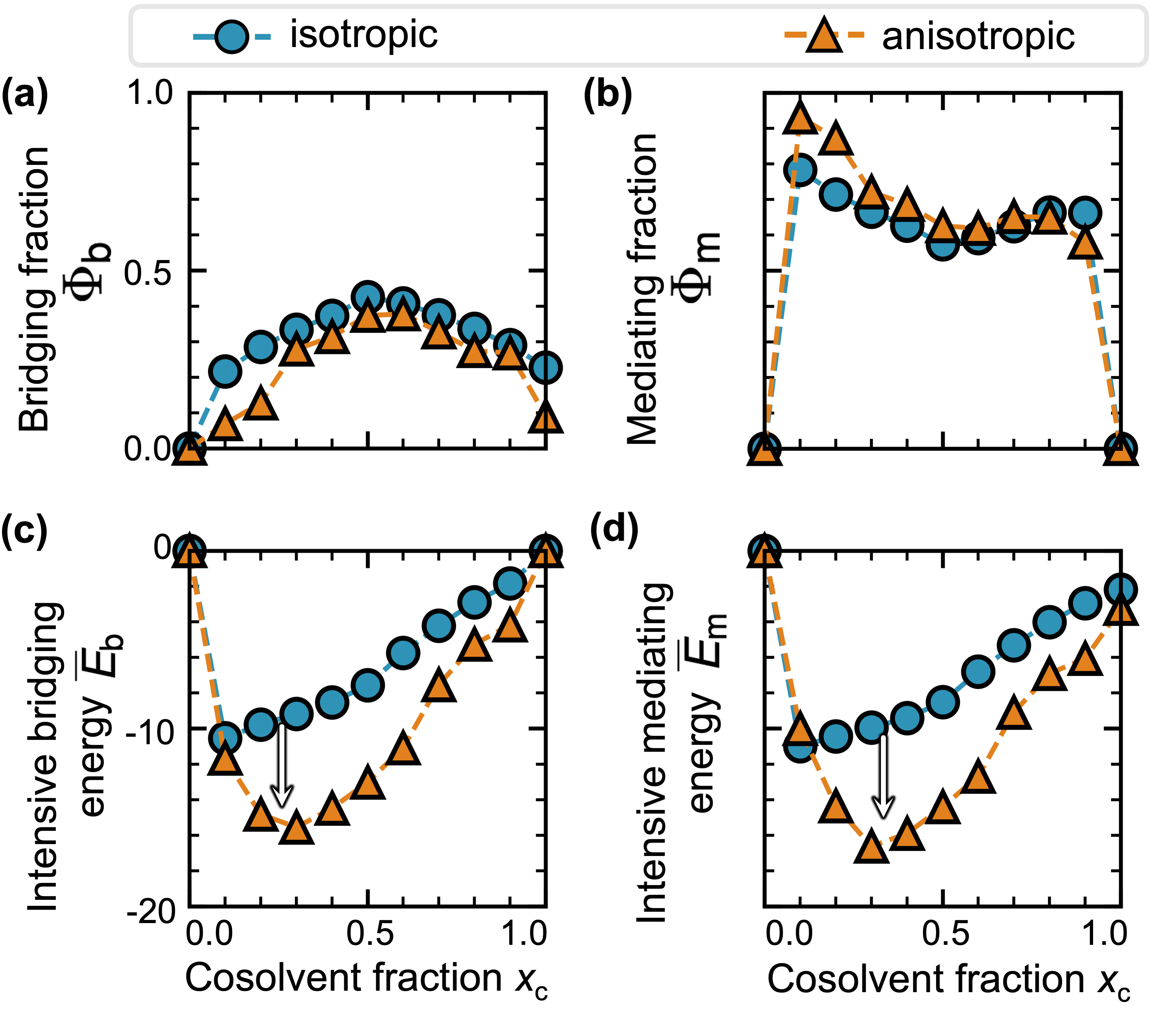}
    \caption{Analysis and comparison of the solvation environment of systems $R_\text{sc}^\circ$ and $R_\text{sc}^\curlywedge$ for $N_\text{m}=32$. (a) Fraction of bridging cosolvents $\Phi _\text{b}$, (b) fraction of mediating cosolvents $\Phi _\text{m}$, (c) energetic contribution per cosolvent particle in the solvation shell $\bar{E}_b$, and (d) the energetic contribution per cosolvent particle in the solvation shell $\bar{E}_m$ as cosolvent fraction $x_c$ is varied from $0$ to $1$. Horizontal axis labels are share betwen panels (a) and (c) as well as (b) and (d).  The white arrows highlight a marked difference in the magnitude of intensive energies. Error bars  correspond to the standard error of the mean and are generally smaller than the symbol size. }
\end{figure}

\begin{figure}[H]
    \centering
    \includegraphics[keepaspectratio]{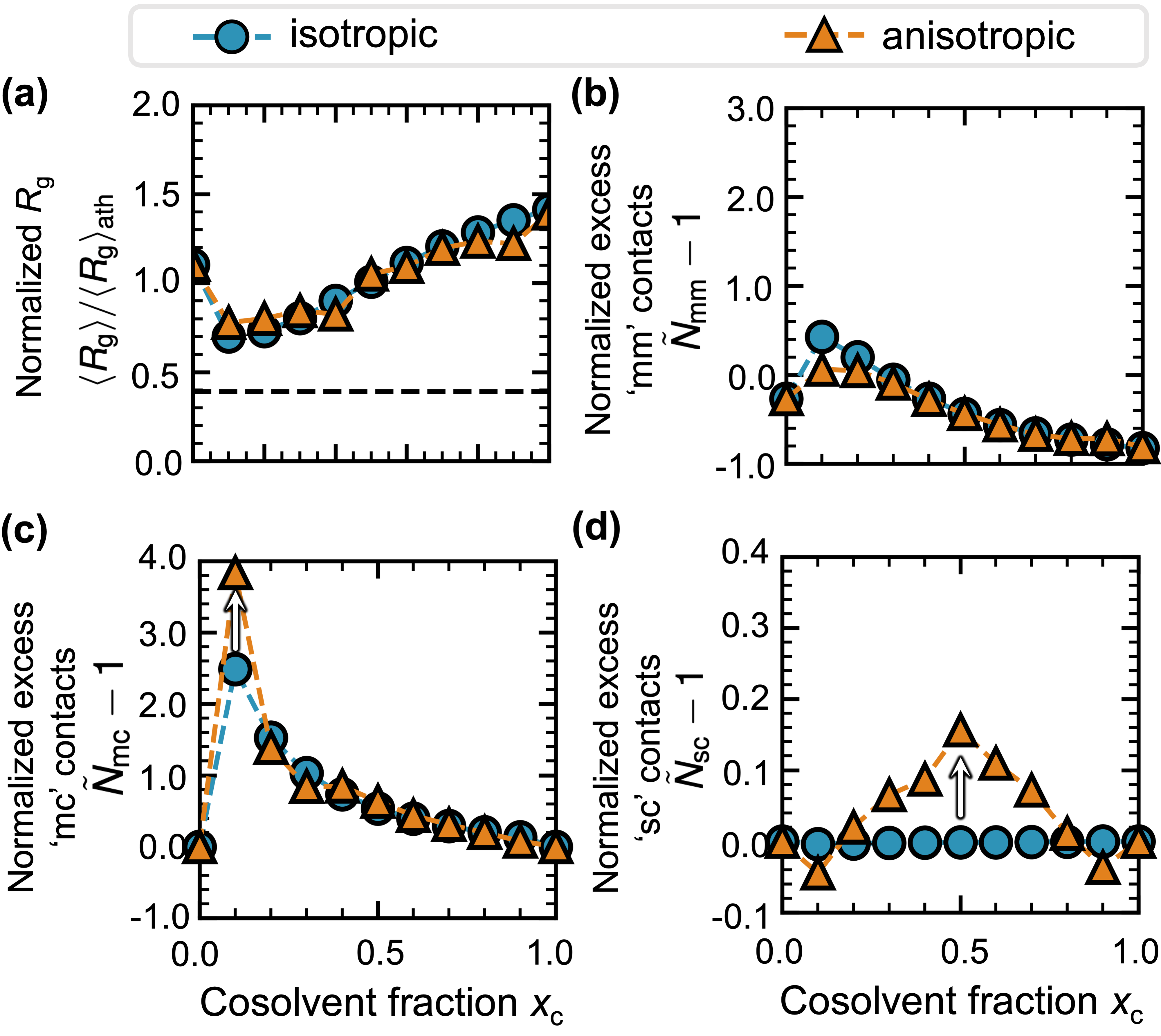}
    \caption{A comparison of conformational characteristics and environment of a polymer from $R_\text{pc}^\circ$ (blue circles) and $R_\text{pc}^\curlywedge$ (orange triangles) for $N_\text{m}=32$. (a) Normalized single-chain $R_\text{g}$ and normalized excess (b) monomer-monomer, (c) solvent-cosolvent, and (d) monomer-cosolvent interactions as cosolvent fraction $x_c$ is varied from $0$ to $1$. In (a), the dashed horizontal line is a guide for the reference to a maximally compact polymer. Horizontal axis labels are share betwen panels (a) and (c) as well as (b) and (d).  In (c) and (d), the white arrow shows a marked shift in the trendlines. Error bars  correspond to the standard error of the mean and are generally smaller than the symbol size. }
\end{figure}

\begin{figure}[H]
    \centering
    \includegraphics[keepaspectratio]{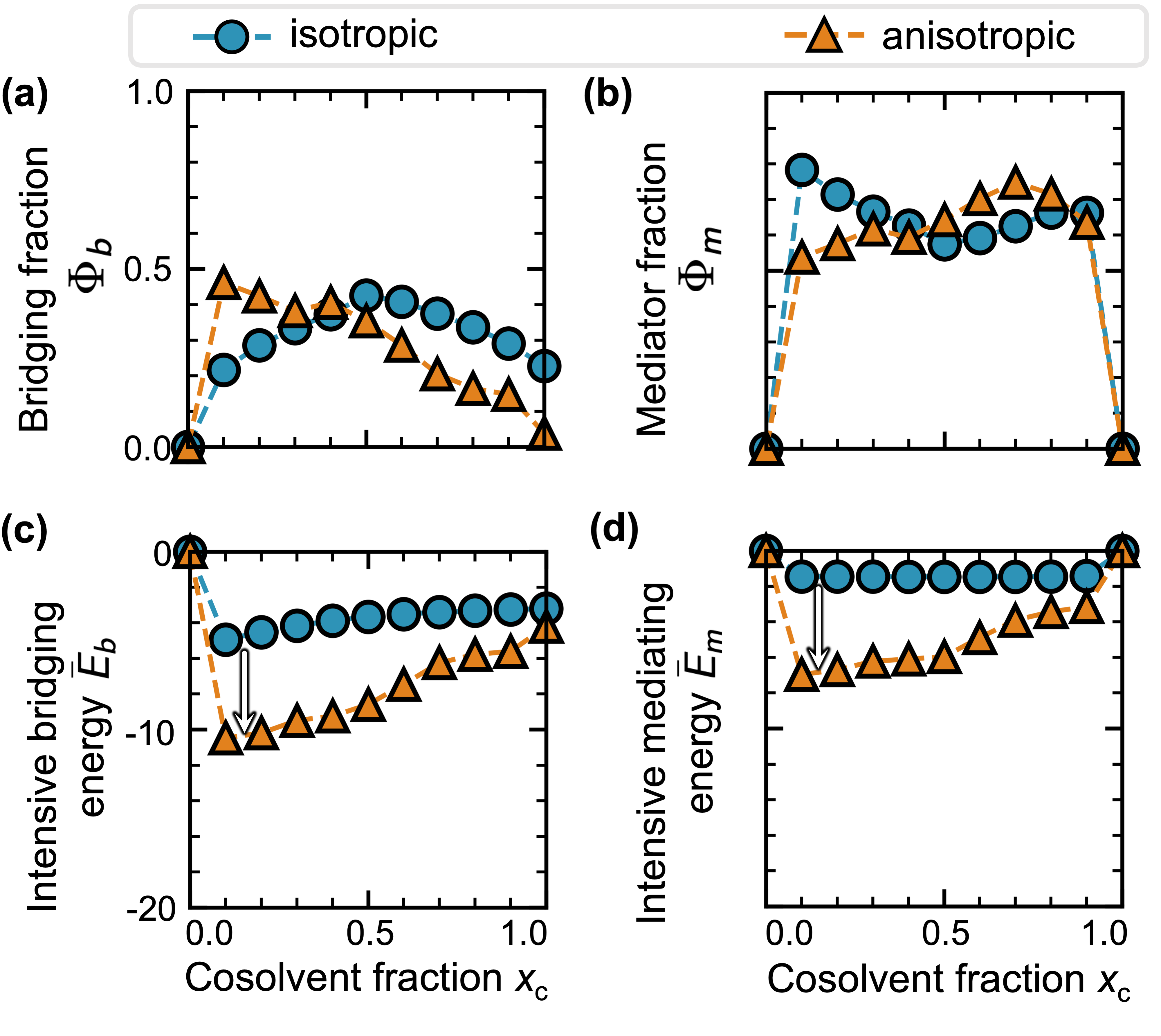}
    \caption{Analysis and comparison of the solvation environment of systems $R_\text{pc}^\circ$ and $R_\text{pc}^\curlywedge$ for $N_\text{m}=32$. (a) Fraction of bridging cosolvents $\Phi _\text{b}$, (b) fraction of mediating cosolvents $\Phi _\text{m}$, (c) energetic contribution per cosolvent particle in the solvation shell $\bar{E}_b$, and (d) the energetic contribution per cosolvent particle in the solvation shell $\bar{E}_m$ as cosolvent fraction $x_c$ is varied from $0$ to $1$. Horizontal axis labels are share betwen panels (a) and (c) as well as (b) and (d).  In (c) and (d), the white arrow shows a marked shift in the intensive energy. Error bars  correspond to the standard error of the mean and are generally smaller than the symbol size.}
\end{figure}